\documentclass[a4paper,11pt]{article}

\usepackage{jheppub} 

\usepackage[T1]{fontenc} 

\title{ Quartic quasi-topological gravity, black holes and holography}

\author[a,b]{M. H. Dehghani}
\author[a]{M. H. Vahidinia}

\affiliation[a]{ Physics Department and Biruni Observatory, College of Sciences, Shiraz
University, Shiraz 71454, Iran}
\affiliation[b]{ Research Institute for Astrophysics and Astronomy of Maragha (RIAAM),
P.O. Box 55134-441, Maragha, Iran}

\emailAdd{mhd@shirazu.ac.ir}
\emailAdd{vahidinia@shirazu.ac.ir}

\abstract{In this paper, we derive the field equations of quartic quasi-topological
gravity by varying the action with respect to the metric. Also, we obtain
the linearized graviton equations in the AdS background and find that it is
governed by a second-order field equation as in the cases of Einstein, Lovelock or
cubic quasi-topological gravities. But in contrast to the cubic quasi-topological gravity,
the linearized field equation around a black hole has fourth-order radial derivative of the perturbation.
Moreover, we analyze the conditions of
having ghost free AdS solutions and AdS planar black holes. In addition, we
compute the central charges of the dual conformal field theory
of this gravity theory by studying holographic Weyl anomaly. Finally, we consider the effect
of quartic term on the causality of dual theory in the tensor channel and show that, in the
contrast to the trivial result of cubic quasi-topological gravity, the
existence of both cubic and quartic terms leads to a non-trivial constraint.
However, this constraint does not imply any lower positive bound on the
viscosity/entropy ratio.}

\begin{document}
\maketitle
\flushbottom

\section{Introduction}

The anti-de Sitter/conformal field theory (AdS/CFT) correspondence provides
an interesting framework to study the non-perturbative quantum field theories.
This kind of duality for the first time introduced in the context of string
theory. However, the most strong version of this conjecture is generalized
to gauge/gravity duality independent of string theory consideration.
According to this duality, in principle, one can perform gravity calculation
to find information about the field theory side or vice versa \cite{AdS}. As
a simple model in the context of AdS/CFT, one can consider Einstein gravity
with a two-derivative bulk action. But in this circumstance, the dual theory
is restricted to the large N and large `t Hooft couplings. Moreover, the
central charges of a conformal field theory (CFT) relate to coupling
constants of its dual gravity. Therefore, Einstein gravity restricts dual
theory to limited class of CFT with equal central charges \cite{CentCharg1}.
In order to extend the duality beyond these limits, one needs to involve
higher curvature terms or derivatives of curvature tensor in the gravity
action. It is clear that each correction term introduces a new coupling
constant and so this procedure leads to the extension of parameter space and
central charges and richness of the CFT theory \cite%
{CentCharg2,FinCoup,Hofman}. It is natural to take the correction terms which
arises from a UV complete theory such as string theory. However, considering
some toy models may help us to learn more about the AdS/CFT duality.

Gauss-Bonnet (second order of Lovelock) term with quadratic interaction or
in general, Lovelock terms in all the orders are well-known corrections
which usually are added to Hilbert-Einstein action. Lovelock gravity has
some interesting features. For example, it leads to second-order field
equation \cite{Lovelock} and so it has ghost free AdS solution \cite%
{GhostFree}. However, because of topological features of Lovelock theory,
Gauss-Bonnet term has no contribution in spacetimes less than
five dimensions and third-order of Lovelock contributes in seven or
higher-dimensional spacetimes. Thus, for studying four-dimensional field
theory with a five-dimensional holographic dual, only Gauss-Bonnet term is
accessible. Throughout the recent years, most of the interesting holographic
aspects of Lovelock gravity, has been studied \cite{Liu,de
Bore1,Edelstein1,Edelstein2,GBd,de Bore2,EdelsteinPaulos,EdelsteinRev}. It
is known that the second order Lovelock gravity (Gauss-Bonnet gravity)
admits supersymmetric extension \cite{Ozkan:2013uk}, while all the
higher-orders of this theory have only the necessary condition of
supersymmetric extension \cite{susy,EdelsteinPaulos}. By these
considerations and for studying the effects of cubic interaction in the
non-supersymmetric gravity on the four-dimensional CFT dual, the cubic
quasi-topological gravity has been constructed \cite{Myers1}. This toy model
provides a second-order field equation for spherically symmetric spacetime
and admits ghost free AdS vacuum and asymptotic AdS black hole solutions in
five dimensions \cite{Myers1,Oliva,Myers2,CubicQuasiBH,CubicQuasiHolo}. As
mentioned in Ref. \cite{Myers2}, one expects that the quasi-topological
gravity may create a better understanding of non-supersymetric holography
and some constraints like causality bound \cite{Liu} and positive energy
restriction \cite{Hofman}. In fact, in the supersymmetric models the positive
energy constraint involves causality bound, so they do not imply independent
bound on the parameters of theory \cite{Hofman,de
Bore1,Edelstein1,Edelstein2,GBd,de Bore2,EdelsteinPaulos,EdelsteinRev}.
Moreover, it has been shown that the causality constraint and positive
energy bound do not match in general, particularly for theories with more
than second derivative in the field equation \cite{Hofman}. Practically,
positive energy and causality constraints are interesting, for example, for
studying the conjecture about the existence of a lower bound on the ratio of
shear viscosity to entropy \cite{Liu,Myers2, Kss}. However, the cubic
quasi-topological gravity interaction does not imply any causality bound
\cite{Myers2}. So the question is: does higher-order quasi-topological
gravity has the same trivial result? By this motivation a more interesting
version of quasi-topological gravity with quartic curvature terms has been
introduced in Ref. \cite{Dehghani1}. In this paper, we are interested in the
studies of the effects of this quartic term on the conditions of existence of
ghost free AdS spacetimes and planar AdS black holes. Moreover, we like
to know some aspects of quartic interaction on the CFT$_{4}$ in context of
AdS$_{5}$/CFT$_{4}$.

Outline of this paper is as follows: We begin with a review on the action of
quartic-topological gravity in Sec. \ref{QT4}. In Sec. \ref{Feq}, we obtain
the general form of fourth-derivatives field equations by varying the action
with respect to the metric $g_{ab}$. Moreover, we use this field equation
for a spherically symmetric spacetime in order to show that the general
fourth-derivative equation reduce to a second-order field equation. Also, we
find the quartic equation which its roots describes the planar AdS solutions
and review the thermodynamics of these black holes. Then in Sec. \ref{Ling},
we show that the first-order perturbation around an AdS solution reduces to
a second-order wave equation for graviton. By using this wave equation, we
obtain the conditions of existence of a ghost free AdS solution. In Sec. %
\ref{Qeq}, we review a few properties of general solutions of quartic
equation. In Sec. \ref{AAdS}, we obtain the constraints on the parameter space of quartic
quasi-topological gravity for having ghost free AdS and planar AdS black
hole. In Sec. \ref{Ano}, we will use the standard
approach of AdS/CFT to compute the central charges of four-dimensional
conformal field theory dual to the five-dimensional quartic quasi-topological
gravity. Section \ref{CC} is devoted to the study of perturbation around a
planar AdS black hole and the possibility of causality violation. Moreover,
the necessary condition for preserving the causality condition in tensor
channel is introduced. Finally in Sec. \ref{Hydro}, we consider the
causality constraint in order to find whether it can imply any positive bound on the
viscosity/entropy ratio. We finish our paper with some concluding remarks.

\section{Review of Quartic Quasi-topological Action in Five Dimensions\label%
{QT4}}

In this section, we review some features of the cubic and quartic
quasi-topological gravity in a five-dimensional spacetime, which may be
considered as the dual of a four-dimensional CFT. This theory of gravity
produces second-order differential equations of motion for a spherical
symmetric spacetime. The bulk gravity action in the presence of a
cosmological constant can be written as:
\begin{equation}
I_{G}=\frac{1}{2l_{p}^{3}}\int d^{5}x\sqrt{-g}[\mathcal{L}_{1}+\frac{\lambda
}{2}L^{2}\mathcal{L}_{2}+\frac{7}{4}\mu L^{4}\mathcal{L}_{3}+\frac{1}{21024}%
\nu L^{6}\mathcal{L}_{4}],  \label{Act}
\end{equation}%
where $\mathcal{L}_{1}=R-2\Lambda $ and $\mathcal{L}_{2}=R_{abcd}{R}^{abcd}-4%
{R}_{ab}{R}^{ab}+{R}^{2}$ are the Einstein-Hilbert and Gauss-Bonnet terms,
and $\mathcal{L}_{3}$ and $\mathcal{L}_{4}$ are the cubic and quartic terms
of quasi-topological gravity, respectively \cite{Myers1,Dehghani1}:
\begin{eqnarray}
\mathcal{L}_{3}
&=&a_{1}R_{a\,\,b}^{\,\,c\,\,\,d}R_{c\,\,d}^{\,\,e\,\,\,f}R_{e\,\,f}^{\,\,a%
\,\,\,b}+a_{2}\,R_{abcd}R^{abcd}R+a_{3}\,R_{abcd}R^{abc}{}_{e}R^{de}  \notag
\\
&&+a_{4}\,R_{abcd}R^{ac}R^{bd}+a_{5}R_{a}{}^{b}R_{b}{}^{c}R_{c}{}^{a}+a_{6}%
\,R_{a}^{\,\,b}R_{b}^{\,\,a}R+a_{7}\,R^{3}  \notag \\
\ \mathcal{L}_{4} &=&c_{1}R_{abcd}R^{cdef}R_{\phantom{hg}{ef}%
}^{hg}R_{hg}{}^{ab}+c_{2}R_{abcd}R^{abcd}R_{ef}R^{ef}+c_{3}RR_{ab}R^{ac}R_{c}{}^{b}+c_{4}(R_{abcd}R^{abcd})^{2}
\notag \\
&&+c_{5}R_{ab}R^{ac}R_{cd}R^{db}+c_{6}RR_{abcd}R^{ac}R^{db}+c_{7}R_{abcd}R^{ac}R^{be}R_{%
\phantom{d}{e}}^{d}+c_{8}R_{abcd}R^{acef}R_{\phantom{b}{e}}^{b}R_{%
\phantom{d}{f}}^{d}  \notag \\
&&+c_{9}R_{abcd}R^{ac}R_{ef}R^{bedf}+c_{10}R^{4}+c_{11}R^{2}R_{abcd}R^{abcd}+c_{12}R^{2}R_{ab}R^{ab}
\notag \\
&&\hspace{-0.1cm}%
+c_{13}R_{abcd}R^{abef}R_{ef}{}_{g}^{c}R^{dg}+c_{14}R_{abcd}R^{aecf}R_{gehf}R^{gbhd},
\label{Lag}
\end{eqnarray}%
\begin{eqnarray}
a_{1} &=&1,\ a_{2}=\frac{3}{8},\ a_{3}=\frac{-9}{7},\ a_{4}=\frac{15}{7},\
a_{5}=\frac{18}{7},\ a_{6}=\frac{-33}{14},\ a_{7}=\frac{15}{56}, \notag\\
c_{1} &=&-1404,\text{ \ \ \ \ }c_{2}=1848,\text{ \ \ \ \ \ }c_{3}=-25536,%
\text{ \ \ \ }c_{4}=-7422,\text{ \ \ \ \ }c_{5}=24672, \notag\\
c_{6} &=&-5472,\text{ \ \ }c_{7}=77184,\text{ \ }c_{8}=-85824{,}\text{ \ \ \
}c_{9}=-41472{,}\text{ \ \ }c_{10}=-690, \notag\\
c_{11} &=&1788,\text{ \ \ }c_{12}=6936,\text{ \ }c_{13}=7296{,}\text{ \ \ }%
c_{14}=42480. \label{cis}
\end{eqnarray}%
As usual we define $\Lambda =-6/L^{2}$, where $L$ is related to $AdS_{5}$
radius.

\section{Field Equation of Quartic Quasi-topological Gravity \label{Feq}}

Here, we will closely follow Ref. \cite{Padd} to derive the field equation
of quasi-topological gravity by varying the action with respect to the
metric $g^{ab}$. Consider a general action
\begin{equation}
I=\int_{\mathcal{M}}d^{n+1}x\sqrt{-g}\left[ \mathcal{L}%
_{g}(g^{ab},R_{bcd}^{a})+\kappa \mathcal{L}_{m}\right],
\end{equation}
where $\mathcal{L}_{m}$ is the matter Lagrangian and $\mathcal{L}%
_{g}(g^{ab},R_{bcd}^{a})$ is a general gravitational Lagrangian, which
contains the metric and Riemann tensors but not any derivatives of $%
R_{bcd}^{a}$. Variation of this action with respect to $g^{ab}$ leads to
\begin{eqnarray}
\delta I &=&\int_{\mathcal{M}}d^{n+1}x\sqrt{-g}(\mathcal{E}_{ab}-\kappa
T_{ab})~\delta g^{ab},  \label{VGAction} \\
\mathcal{E}_{ab} &=&\frac{1}{\sqrt{-g}}\frac{\partial \sqrt{-g}\mathcal{L}%
_{g}}{\partial g^{ab}}-2\nabla ^{m}\nabla ^{n}P_{amnb}~;~\ \ \ \ \ \
P_{a}^{~\ bcd}\equiv \frac{\partial \mathcal{L}_{g}}{\partial R_{\ bcd}^{a}},
\end{eqnarray}
where $T_{ab}$ is the energy momentum tensor of matter field. By considering
$\partial \sqrt{-g}/\partial g^{ab}=-\sqrt{-g}g_{ab}/2$, one obtains
\begin{eqnarray}
\frac{1}{\sqrt{-g}}\frac{\partial \sqrt{-g}\mathcal{L}_{g}}{\partial g^{ab}}
&=&\frac{\partial \mathcal{L}_{g}}{\partial R_{ij}^{kl}}\frac{\partial
R_{ij}^{kl}}{\partial g^{ab}}-\frac{1}{2}g_{ab}~\mathcal{L}_{g}  \notag \\
&=&P_{b}^{\ \ \ kij}R_{akij}-\frac{1}{2}g_{ab}~\mathcal{L}_{g}.
\end{eqnarray}
So, the field equation can be written as
\begin{equation}
P_{b}^{\ kij}R_{akij}\mathcal{-}2\nabla ^{m}\nabla ^{n}P_{amnb}-\frac{1}{2}%
g_{ab}~\mathcal{L}_{g}=\kappa T_{ab}.  \label{Geq}
\end{equation}
To have an explicit form of the field equation, one needs to calculate $%
P_{abcd~\ }$in term of Riemann and metric tensor. In general, the field
equation (\ref{Geq}) contains fourth-order derivatives of the metric due to
the existence of $\nabla ^{m}\nabla ^{n}P_{amnb}$. But for Einstein and any
Lanczos-Lovelock Lagrangian $\nabla ^{n}P_{amnb}=0$, and therefore the field
equation (\ref{Geq}) contains at most second-order derivatives of the metric.

Using standard software \cite{Cad}, it is a matter of calculation to show
that $P_{abcd}$\ for the Lagrangian of quasi-topological gravity (\ref{Lag})
is
\begin{equation}
P_{a}^{\ bcd}=P_{(1)a}^{\ \ \ \ bcd}+\frac{\lambda }{2}L^{2}P_{(2)a}^{\ \ \
\ bcd}+\frac{7}{4}\mu L^{4}P_{(3)a}^{\ \ \ \ bcd}+\frac{1}{21024}\nu
L^{6}P_{(4)a}^{\ \ \ \ bcd},  \label{Pabcd}
\end{equation}%
where
\begin{equation}
P_{(i)abcd}^{\ \ \ \ }=\frac{1}{2}(\Pi _{(i)[ab][cd]}+\Pi _{(i)[cd][ab]}).
\label{Pabcd2}
\end{equation}%
In Eq. (\ref{Pabcd2}), $\Pi _{(i)abcd}$'s are
\begin{equation}
{\Pi }_{(1)a}\,^{bcd}=g_{a}^{c}g^{bd},\text{ \ \ \ }{\Pi }%
_{(2)a}\,^{bcd}=2R_{a}^{\ bcd}-8g_{a}^{c}R^{bd}+2g_{a}^{c}g^{bd}R,
\end{equation}%
\begin{eqnarray}
{\Pi }_{(3)a}\,^{bcd} &=&a_{1}(3R^{bmdn}R_{man}^{\ \ \ \ \ \ \
c})+a_{2}(2RR_{a}^{\ \ bcd}+R_{mnop}R^{mnop}g_{a}^{c}g^{bd})  \notag \\
&&+a_{3}(2R_{a}^{\ \ bcm}R_{m}^{d}+R^{mnod}R_{mno}^{\ \ \ \ \ \ \
b}g_{a}^{c})+a_{4}(2R^{bmdn}R_{mn}g_{a}^{c}+R_{a}^{c}R^{bd})  \notag \\
&&+a_{5}(3R_{a}^{m}R_{m}^{c}g^{bd})+a_{6}(2RR_{a}^{c}g^{bd}+R_{mn}R^{mn}g_{a}^{c}g^{bd})+a_{7}(3R^{2}g_{a}^{c}g^{bd}),
\end{eqnarray}%
\begin{eqnarray}
{\Pi }_{(4)a}\,^{bcd} &=&c_{1}(2{R}^{mn}\,_{a}\,^{b}{R}_{mn}\,^{op}{R}%
^{cd}\,_{op}+2{R}^{mnop}{R}_{a}\,^{b}\,_{mn}{R}_{op}\,^{cd})  \notag \\
&&+c_{2}(2\,{R}_{a}\,^{bcd}{R}_{mn}{R}^{mn}+{2R}^{bd}{R}^{mnop}{R}_{mnop}{g}%
_{a}\,^{c}),  \notag \\
&&+c_{3}(R^{mn}R_{mo}R_{n}^{o}g_{a}^{c}g^{bd}+R_{am}R^{cm}Rg^{bd}+2g_{a}^{c}R^{bm}R_{m}^{d}R)+c_{4}(4%
{R}_{a}\,^{bcd}{R}^{mnop}{R}_{mnop})  \notag \\
&&+c_{5}(4\,{R}^{bm}{R}^{dn}{R}_{mn}{g}_{a}\,^{c})+c_{6}({R}^{no}{R}^{qr}{R}%
_{nqor}{g}_{a}\,^{c}{g}^{bd}+R{R}_{a}\,^{c}{R}^{bd}+2R{R}^{no}{R}%
^{b}\,_{n}\,^{d}\,_{o}{g}_{a}\,^{c})  \notag \\
&&+\,c_{7}({R}_{a}\,^{c}{R}^{bm}{R}^{d}\,_{m}+{R}^{mn}{R}_{m}\,^{p}{R}%
^{b}\,_{n}\,^{d}\,_{p}{g}_{a}\,^{c}+{R}^{dm}{R}^{op}{R}_{o}\,^{b}\,_{pm}{g}%
_{a}\,^{c}+{R}_{a}\,^{n}{R}^{pq}{R}_{pnq}\,^{c}{g}^{bd})  \notag \\
&&+c_{8}({R}^{bm}{R}^{dn}{R}_{a}\,^{c}\,_{mn}+{R}^{cm}{R}^{dn}{R}%
_{am}\,^{b}\,_{n}+{R}^{mn}{R}_{am}\,^{op}{R}_{o}\,^{c}\,_{pn}{g}^{bd}-{R}%
^{mn}{R}^{o}\,_{m}\,^{pc}{R}_{anop}{g}^{bd})  \notag \\
&&+c_{9}({R}_{a}\,^{c}{R}^{mn}{R}^{b}\,_{m}\,^{d}\,_{n}+2{R}^{mn}{R}%
_{m}\,^{o}\,_{n}\,^{p}{R}^{b}\,_{o}\,^{d}\,_{p}{g}_{a}\,^{c}+{R}^{bd}{R}^{no}%
{R}_{nao}\,^{c})+c_{10}(4R^{3}{g}_{a}\,^{c}{g}^{bd})  \notag \\
&&+c_{11}(2\,R{R}^{mnop}{R}_{mnop}{g}_{a}\,^{c}{g}^{bd}+2\,R^{2}{R}%
_{a}\,^{bcd})+c_{12}(2\,R{R}^{mn}{R}_{mn}{g}_{a}\,^{c}{g}^{bd}+2\,R^{2}{R}%
^{bd}{g}_{a}\,^{c})  \notag \\
&&+c_{13}(2{R}^{dm}{R}^{npc}\,_{m}{R}_{a}\,^{b}\,_{np}+{R}^{mn}{R}%
_{a}\,^{bp}\,_{m}{R}^{cd}\,_{pn}+{R}^{mnob}{R}^{pq}\,_{o}\,^{d}{R}_{mnpq}{g}%
_{a}\,^{c})  \notag \\
&&+c_{14}(4\,{R}_{a}\,^{mcn}{R}^{obpd}{R}_{monp}).
\end{eqnarray}%
One can show that $\nabla ^{n}P_{amnb}=0$ for static \footnote{Also, $\nabla ^{n}P_{amnb}=0$ for Friedmann-Robertson-Walker metric in any dimensions
in the quartic quasi-topological gravity.}
spacetimes (with flat, spherical or hyperbolic boundary) in the quartic quasi-topological gravity, and therefore the field equation is at most second-order
differential equation.

\subsection{Planar AdS Black Holes in 5 dimensions}

The metric of a five-dimensional planar AdS static spacetime
may be written as
\begin{equation}
ds^{2}=\frac{r^{2}}{L^{2}}(-N^{2}(r)f(r)dt^{2}+dx^{2}+dy^{2}+dz^{2})+\frac{%
L^{2}}{r^{2}f(r)}dr^{2}.  \label{metric}
\end{equation}%
Using the field equations of quasi-topological gravity, one obtains
\begin{gather}
\frac{3}{2L^{4}r}N(r)^{2}f(r)[r^{4}(1-f+\lambda f^{2}+\mu f^{3}+\nu
f^{4})]^{\prime }=0, \\
\frac{3}{2r^{5}f(r)}[r^{4}(1-f+\lambda f^{2}+\mu f^{3}+\nu f^{4})]^{\prime }
\notag \\
-\frac{3}{rN(r)}[1-2\lambda f-3\mu f^{2}-4\nu f^{3}]N(r)^{\prime }=0.
\end{gather}%
So, the metric functions must satisfy the following equations
\begin{gather}
\gamma (f)\equiv 1-f(r)+\lambda f^{2}(r)+\mu f^{3}(r)+\nu f^{4}(r)-\frac{%
r_{h}^{4}}{r^{4}}=0,  \label{f(r)} \\
N^{\prime }(r)=0,
\end{gather}%
where the integration constant $r_{h}$ is the radius of black hole horizon.
Moreover, the last equation shows that $N(r)$ is a constant which we set it $%
N(r)=f_{\infty }^{-1/2}$ to fix the speed of light on the boundary equal to one. The constant $f_{\infty }$ is determined by taking $%
r\rightarrow \infty $ (or equally $r_{h}\rightarrow 0$) limit of Eq. (\ref%
{f(r)}). Thus, it is one of the roots of the following quartic equation
\begin{equation}
\gamma (f_{\infty })\equiv 1-f_{\infty }+\lambda f_{\infty }^{2}+\mu
f_{\infty }^{3}+\nu f_{\infty }^{4}=0.  \label{finf roots}
\end{equation}%
One may note that the radius of the AdS vacuum solution is $L_{eff}=Lf_{\infty }^{-1/2}$
\footnote{It is worth to mention that
in a D-dimensional ($D>8$) isotropic spacetime with constant curvature boundary, the field equation of quasi-topological gravity
is the same as that of quartic Lovelock theory. Thus, the black holes of quartic quasi-topological gravity are the
same as those of quartic Lovelock gravity in $D>8$ dimensions. Detailed analysis
of non-planar black holes of quartic Lovelock gravity are provided in \cite{Lovebes}. Of course, the cubic and the quartic terms of Lovelock gravity vanish for $D<7$ and $D<9$, respectively.}.

The black hole solutions of Eq. (\ref{f(r)}) have a horizon at $r=r_{h}$.
The Hawking temperature of this horizon is given by
\begin{equation}
T=\frac{r_{h}}{\pi L^{2}\sqrt{f_{\infty }}}.  \label{Temper}
\end{equation}
Moreover, the energy and entropy densities are
\begin{equation}
M=\frac{3r_{h}^{4}}{2l_{p}^{3}L^{5}\sqrt{f_{\infty }}},\ \ \ \ \ \ \ \ \ \ \
\ \ \ s=\frac{2\pi r_{h}^{3}}{l_{p}^{3}L^{3}},
\end{equation}
and therefore as like as any natural 5-dimensional black hole duals to
a thermal CFT$_{4}$, the relation between these quantities are $M=3Ts/4$.

\section{Linearized Graviton Equation in the AdS$_{5}$ Vacuum \label{Ling}}

As we mentioned, the field equation of quasi-topological gravity for
static spacetime is second-order differential equations. But,
here we like to study the field equation under a small perturbation around
the AdS$_{5}$ background. For cubic quasi-topological gravity, the authors
of Ref. \cite{Myers1} examined the linearized equations of motion for a
graviton perturbation, and showed that the linearized graviton equation in
an AdS background is second-order. Here, we like to do the same job in
quartic quasi-topological gravity. We examine the following trial
perturbation around the AdS$_{5}$ background:
\begin{equation}
ds^{2}=\frac{r^{2}}{L^{2}}(-{dt^{2}}+dx^{2}+dy^{2}+dz^{2})+\frac{L^{2}}{%
f_{\infty }}\frac{dr^{2}}{r^{2}}+2\varepsilon h(t,r,x,z)dxdy+2\varepsilon
k(t,r,x,z)dtdr.  \label{pert metric}
\end{equation}%
Using Eq. (\ref{Pabcd}), one finds that $\nabla
^{m}\nabla ^{n}P_{amnb}=0+O(\varepsilon ^{2})$. For instance, the graviton
equations $\mathcal{E}_{rt}$ and $\mathcal{E}_{xy}$ in the AdS background up
to first order in $\varepsilon $ are
\begin{gather}
-\frac{\varepsilon }{2}f_{\infty }(1-2\lambda f_{\infty }-3\mu f_{\infty
}^{2}-4\nu f_{\infty }^{3})\frac{L^{2}}{r^{2}}[\frac{\partial ^{2}}{\partial
x^{2}}h(t,r,x,z)+\frac{\partial ^{2}}{\partial z^{2}}k(t,r,x,z)]=\kappa
T_{rt},  \label{graviton eq} \\
-\frac{\varepsilon }{2}(1-2\lambda f_{\infty }-3\mu f_{\infty }^{2}-4\nu
f_{\infty }^{3})\frac{1}{L^{2}r^{2}}[-L^{4}\frac{\partial ^{2}}{\partial
t^{2}}+r^{4}f_{\infty }\frac{\partial ^{2}}{\partial r^{2}}  \notag \\
+r^{3}f_{\infty }\frac{\partial }{\partial r}+L^{4}\frac{\partial ^{2}}{%
\partial z^{2}}-4r^{2}f_{\infty }]h(t,r,x,z)=\kappa T_{xy},  \label{GravEq2}
\end{gather}%
which show that the propagation of a graviton in an AdS background is
governed by a second-order equation. $T_{ab}$, which has been added to the
right-hand side of the field equation, is due to the minimally coupling of
the metric and the matter field which creates the perturbation, or it can be
taken into account for the higher order contributions in the graviton. Here,
we should emphasize that we have tried various small perturbation around AdS$%
_{5}$ background to make sure that the above result is true for any first
order perturbation. It seems that the above equations for a general small perturbation
may be written in the same form as that of cubic quasi-topological gravity \cite{Myers1,Myers2}. That is
\begin{gather}
-\frac{1}{2}(1-2\lambda f_{\infty }-3\mu f_{\infty }^{2}-4\nu f_{\infty
}^{3})[\nabla ^{2}h_{ab}+\nabla _{a}\nabla _{b}h_{c}^{c}-\nabla _{a}\nabla
^{c}h_{cb}-\nabla _{b}\nabla ^{c}h_{ac}  \notag \\
+g_{(0)ab}(\nabla ^{c}\nabla ^{d}h_{cd}-\nabla ^{2}h_{c}^{c})+\frac{2}{%
L_{eff}^{2}}(h_{ab}-g_{(0)ab}h_{c}^{c})]=\kappa T_{ab},
\end{gather}%
where $\nabla _{a}$ is the covariant derivative of the AdS$_{5}$ metric and $%
h_{ab}$ is the perturbation around the AdS$_{5}$ background $g_{(0)ab}$.
This result up to an overall constant factor is the same as the propagation
of a graviton in the AdS$_{5}$ background of the Einstein's theory of
gravity ($\lambda =\mu =\nu =0$). However, one can think about the sign of
this overall factor and\ compare it with Einstein case, to conclude that one
has a ghost free theory (as like as Einstein theory) provided
\begin{equation}
-\gamma ^{\prime }(f_{\infty })=1-2\lambda f_{\infty }-3\mu f_{\infty
}^{2}-4\nu f_{\infty }^{3}>0.  \label{ghost free}
\end{equation}%
If this condition is violated, the kinetic term in the action will be
appeared with an opposite common sign. Moreover, we will see in the
following sections that how this property relates to the positive mass
assumption of the black hole solution and unitary condition of dual theory.
The fact that gravitons propagating in an AdS background simply obey the
same equation of motion as in the Einstein gravity plays an important role
in the understanding of the holographic properties of cubic quasi-topological gravity \cite{Myers2}
and quartic quasi-topological gravity introduced in Ref. \cite{Dehghani1}.

\section{General Solutions of Quartic Equations \label{Qeq}}

The most general form of a quartic equation is given by
\begin{equation}
f^{4}+af^{3}+bf^{2}+cf+d=0,
\end{equation}%
which admits four solutions as
\begin{eqnarray*}
f_{1} &=&\frac{1}{4}\left[-a-K-\sqrt{-(H+K^{2})+2\frac{G}{K}}\right],\ f_{2}=\frac{1}{4}%
\left[-a-K+\sqrt{-(H+K^{2})+2\frac{G}{K}}\right], \\
f_{3} &=&\frac{1}{4}\left[-a+K-\sqrt{-(H+K^{2})-2\frac{G}{K}}\right],\ f_{4}=\frac{1}{4}%
\left[-a+K+\sqrt{-(H+K^{2})-2\frac{G}{K}}\right],
\end{eqnarray*}%
where
\begin{eqnarray*}
&&K=\left[\frac{2^{7/3}(I+2^{-2/3}(\sqrt{-\Delta }-J)^{2/3})}{3(\sqrt{-\Delta }%
-J)^{1/3}}-\frac{H}{3}\right]^{1/2}, \\
&&I=b^{2}+12d-3ac,~\ J=9b(ac+8d)-27(c^{2}+a^{2}d)-2b^{3}, \\
&&G=a^{3}-4ab+8c,\ \ H=8b-3a^{2},\text{ \ \ }\Delta =4I^{3}-J^{2}.
\end{eqnarray*}%
As one may see in Ref. \cite{Math}, the general reality conditions of these
solutions can be determined by the signs of $G$, $H$, $\Delta $ and
\begin{equation*}
\delta =3a^{4}-16a^{2}b+16ac+16(b^{2}-4d).
\end{equation*}%
The nature of roots is as follows:

\textbf{I.} Two real and two complex roots: If $\Delta <0$, then there are
two real and two complex roots.

\textbf{II.} Four real roots: If $\Delta >0,~\delta >0,~H<0$ , then all the
four roots are real.

\textbf{III.} No real roots: If $\Delta >0,~\delta <0,~H<0~$or $\Delta
>0,~H\geq 0$, then there exists no real root. In order to \ find the
complete analysis for the special cases where either of $\Delta $, $\delta $,%
$\ H$ or $G$ vanishes one can see \cite{Math}.

\section{Asymptotic AdS Black Hole Solutions \label{AAdS}}

In this section, we study the conditions of existence of asymptotic AdS
black hole solutions. Indeed, as we will see, the AdS spacetime and AdS
black hole exist only in a limited region of the full parameter space.

\subsection{Ghost Free AdS Solutions}

As we mentioned before, the asymptotic AdS solution exists when the large $r$
limit of the quartic equation $\gamma (f)\equiv 1-f+\lambda f^{2}+\mu
f^{3}+\nu f^{4}-r_{h}^{4}/r^{4}=0$ has positive real roots. That is, the
quartic equation (\ref{finf roots}) should have positive real root. Indeed,
the cases with $f_{\infty }<0$ relate to dS solutions. So, we should find
the conditions on the parameters $\lambda ,$ $\mu $ and $\nu $ that Eq. (\ref%
{finf roots}) admits positive real roots. Moreover, we are interested in
ghost free theory and therefore these real positive roots should fulfil
the ghost free condition (\ref{ghost free}). To see the effects of
quasi-topological terms, we first limit our study in the absence of Gauss-Bonnet term ($\lambda =0$), and give some comments in the presence of Gauss-Bonnet term ($\lambda \neq 0$) at the end of this section.
In the absence of Gauss-Bonnet term, the discriminant functions $\Delta ,~\delta $, $H$ and $G$ reduce to
\begin{eqnarray}
\Delta &=&[108(4\nu +\mu )^{3}-729(\nu +\mu ^{2})^{2}]\nu ^{-6},  \notag \\
\delta &=&\frac{3\mu ^{4}}{\nu ^{4}}-\frac{16\mu }{\nu ^{2}}-\frac{64}{\nu }%
,\ \ H=\frac{-3\mu ^{2}}{\nu ^{2}}<0,\text{ \ }G=\frac{\mu ^{3}}{\nu ^{3}}-%
\frac{8}{\nu }.
\end{eqnarray}%
Using the fact that $\gamma (0)=1$ and the large-$x$ behavior of $\gamma
(x)=1-x+\mu x^{3}+\nu x^{4}$, one can analyze the roots of $\gamma
(f_{\infty })=0\ $as follow. For $x\rightarrow \pm \infty $, the quartic
term dominates, $\gamma (x)\approx \nu x^{4}$, and therefore the sign of $%
\nu $ determines the behavior of polynomials $\gamma (x)$ for asymptotic
values $x\rightarrow \pm \infty $ . In addition, $H<0$. Thus, the analysis
of the solutions is as follows:

\textbf{I.} For the case (I) discussed in previous section with two real and
two complex roots, two cases happen.

\textbf{a.}$~\nu >0$: in this case the Vieta's formulas $\Pi
_{i=1}^{4}f_{\infty (i)}=1/\nu $ implies that the two real roots have the
same signs. This is due to the fact that the two complex roots are conjugate
and their product are positive. Therefore in this case, there are two AdS or
two dS vacua where always one of the AdS solution is ghosty and the other
is ghost free. However, one can check that for positive $\mu $, when $G<0$
the two roots are positive and for $G>0$ the two roots are negative.
Moreover, for $\mu <0$ the roots are positive \cite{Math}. In summary, there
are one ghosty AdS and one ghost free AdS for $\mu <0$ and for $\mu >0$ with negative $G$.

\textbf{b. }$\nu <0$: by using the Vieta's formulas, it is easy to see there
are one AdS and one dS solution. In this case by considering the slope of $%
\gamma ^{\prime }(f_{\infty })<0$ and the asymptotic behavior $f_{\infty
}\rightarrow \pm \infty $, one finds that the AdS solution is ghost free.

\begin{figure}[tbp]
\centering
\includegraphics[width=0.6\textwidth]{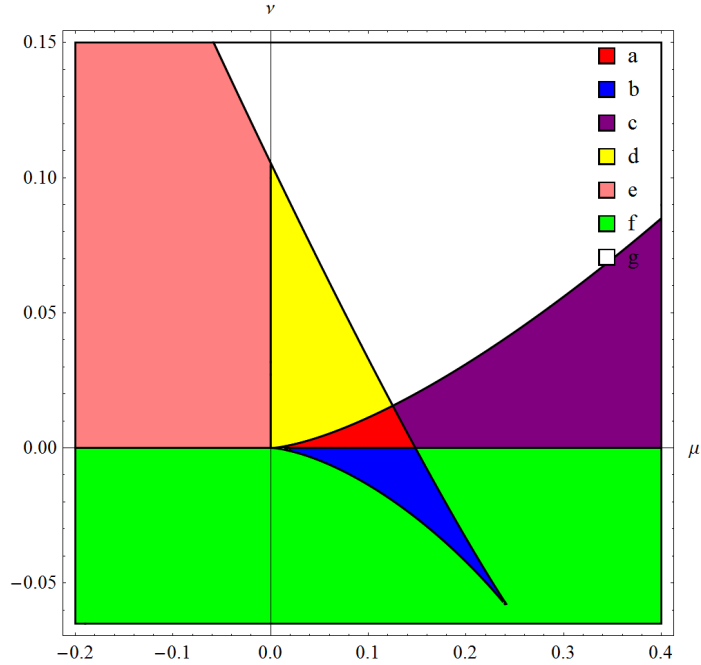}
\caption{Regions (a) to (g) displayed in the parameter space $(\protect\mu ,%
\protect\nu )$. There is no ghost free AdS in region (c). Moreover, the
region (g) does not admit any AdS solution. See table \protect\ref{Tab1} for
more details about the other regions.}
\label{munu}
\end{figure}

\textbf{II.} For case (II) discussed in the previous section with four real
roots, one may find from Fig. \ref{munu} that this case happens only when $%
\mu >0$. Again, two cases happen.

\textbf{a}. $\nu >0$: by using Vieta's formulas $f_{\infty (1)}f_{\infty
(2)}+f_{\infty (1)}f_{\infty (3)}+...+f_{\infty (3)}f_{\infty (4)}=\lambda
/\nu =0$, one may see that it is not possible that all the roots have the
same sign. In addition, $\gamma (0)=1$ and the asymptotic behavior $\gamma
(f_{\infty })\approx \nu f_{\infty }^{4}~$ show there are two positive and
two negative roots. So, in this case there are two dS solutions and two AdS
where only one of the AdS solution is ghost free.

\textbf{b. }$\nu <0$: the Vieta's formulas. $\Pi _{i=1}^{4}f_{\infty
(i)}=1/\nu $ shows that the numbers of negative roots are odd. But, it is a
matter of numerical analysis to show that it is not possible to have three
negative roots. Thus there are three AdS
solutions while only one of them is ghost free.

\subsection{Ghost Free AdS Planar Black Holes}

Now, we study the conditions of existence of AdS planar black holes. In order
to have this kind of black hole, the metric function $f(r)$ must have a few
properties. First, one may expect for large $r$ the metric function behaves
as
\begin{equation}
f(r)=f_{\infty }-\frac{m}{r^{4}}+O(\frac{1}{r^{5}}),  \label{Asym f(r)}
\end{equation}
where $m$ is related to the mass of black hole and should be positive. By
substituting this expansion in Eq. (\ref{f(r)}) and using Eq. (\ref{finf
roots}), one obtains
\begin{equation}
m=\frac{\gamma (f_{\infty })}{\gamma ^{\prime }(f_{\infty })}r^{4}-\frac{%
r_{h}^{4}}{\gamma ^{\prime }(f_{\infty })},
\end{equation}
which shows that for a ghost free ($\gamma ^{\prime }(f_{\infty })<0$) AdS
black hole solution ($\gamma (f_{\infty })=0$), $m$ is positive. This fact
relates the problem of negative mass to the problem of ghost in the AdS
background.

\begin{figure}[tbp]
\centering
\includegraphics[width=0.7\textwidth]{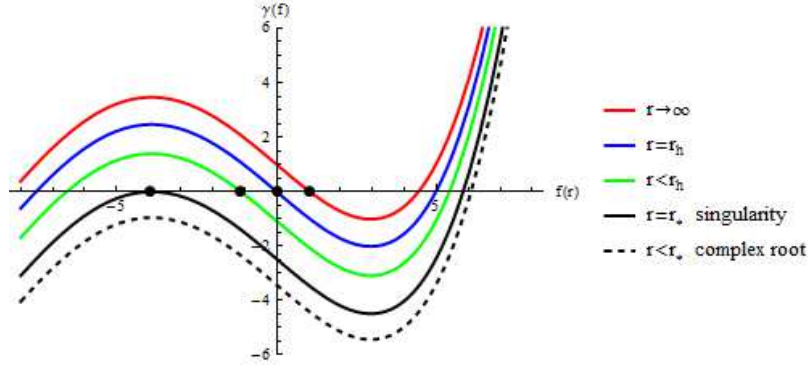}
\caption{$\protect\gamma (f)$ versus $f(r)$ for different $r$. Black dots show the value of $f(r)$ for a black hole solution
at different $r$. Note how $f(r)$ increases from a negative value at $r=r_{\ast}$ to a
positive value as $r\to \infty$, and it is complex for $r<r_{\ast}$.}
\label{comproot}
\end{figure}

Second, for an AdS black hole solution $f(r)$ must be a real monotonous
increasing function in the domain $r\in (r_{\ast },\infty )$ from the
negative value $f_{\ast }=f(r_{\ast })$ to $f_{\infty }>0$ (see Fig. (\ref{comproot})) \footnote{However, for some cases $r_{\ast }=0$}. As we will show
later, the spacetime is singular at $r_{\ast }$ and $f(r)$ is complex for
the excluded region $r<r_{\ast }$ (see Fig. \ref{comproot}). In this case, $%
f(r)$ is zero at $r=r_{h}$ and therefore $f(r)$ describes a black hole
solution. This shows that $f_{\infty }$ in Eq. (\ref{Asym f(r)}) must be the
smallest positive real root of $\gamma (f_{\infty })=0$. This requirement
excludes any extremum for $\gamma =\gamma (x)$ between $x=0~$and the
smallest positive real root $x=f_{\infty }$ \cite{de Bore2}. Since, any
extremum in this domain indicates the possibility of a naked singularity
\cite{Myers1,Takahashi}. To be more clear, suppose that there is an extremum
for $\gamma (x)$. One may note that any extremum of $\gamma (x)$\ related to
a double root solution of $\gamma (f)$\ at $r=r_{\ast }$. Now, expanding $%
\gamma (f)$\ around $\ r=r_{\ast }$, one obtains
\begin{equation*}
f(r)\simeq f_{\ast }-\frac{\gamma ^{\prime }}{2\gamma ^{\prime \prime }}\pm
\frac{\alpha }{\gamma ^{\prime \prime }r_{\ast }^{3}}\sqrt{\beta _{1}\gamma
^{\prime }+(\beta _{2}\gamma ^{\prime }+\beta _{3})(r-r_{\ast })}
\end{equation*}%
where $f_{\ast }=f(r_{\ast })$ and $\alpha $ and $\beta _{i}$'s are constant
factors. So,~one can show that the behavior of the Kretchman scalar at $%
r=r_{\ast }$ in term of $\gamma ^{\prime }$ is
\begin{equation*}
R_{abcd}R^{abcd}\propto \frac{1}{\gamma ^{\prime 3}}+O(\frac{1}{\gamma
^{\prime ^{2}}}).
\end{equation*}%
However, a double root occurs when $\gamma (f_{\ast })=0=\gamma ^{\prime
}(f_{\ast })$, and therefore we have an essential singularity at $r=r_{\ast
} $. But, one may note that any maximum for $x<0$ may relate to a
singularity behind a horizon, in this case $f(r)$ is complex for excluded
region $r<r_{\ast }$ (see Fig. \ref{comproot}). In summary, there is a black
hole solution if $\gamma (f_{\infty })=0$ has a positive real root with $%
\gamma ^{\prime }(f_{\infty })<0$ and $\gamma (f)$ decreases monotonously
in domain $(0,f_{\infty })$.

In order to analyze the extrema of $\gamma =\gamma (x)$, we may study the
roots of $\gamma ^{\prime }(\tilde{x})=0$,
\begin{equation}
\tilde{x}^{3}+\frac{3\mu }{4\nu }\tilde{x}^{2}-\frac{1}{4\nu }=0.
\end{equation}%
The Vieta's formulas $\tilde{x}_{1}\tilde{x}_{2}\tilde{x}_{3}=(-1)^{3}(-1/4%
\nu )$ and $\tilde{x}_{1}\tilde{x}_{2}+\tilde{x}_{1}\tilde{x}_{3}+\tilde{x}%
_{2}\tilde{x}_{3}=0$ imply
\begin{equation}
\frac{1}{\tilde{x}_{1}}+\frac{1}{\tilde{x}_{2}}+\frac{1}{\tilde{x}_{3}}=0.
\label{extrerma}
\end{equation}%
Moreover, a cubic equation like $\gamma ^{\prime }(\tilde{x})=0$ has three
real roots or one real and two complex roots. The former leads to the
possibility of the existence of two negative roots and one positive or vice
versa. The later shows the real root may be either positive or negative. On
the other hand, The sign of product of roots depends on sign of $\nu $. By
using these hints, one finds that for $\nu >0$ there are one real positive
and two complex roots or one positive and two negative real roots. Also for $%
\nu <0,~$there are one real negative and two complex roots or one negative
and two positive real roots. In addition, one may consider the condition of
the existence of complex roots for our cubic equation by investigating the
sign of the discriminant function $\mathcal{D}=4\nu ^{4}-\mu ^{3}\nu ^{2}.$
When $\mathcal{D}>0$ there are one real and two complex roots and for $%
\mathcal{D}<0~$all the roots are real. We will ignore special cases where $%
\mathcal{D}=0$ here (see \cite{Myers1} for useful information on cubic
equation). By using these facts, we may easily study the conditions of
existence of black hole as follows (see table \ref{Tab1}).

\begin{figure}[tbp]
\centering
\includegraphics[width=0.35\textwidth]{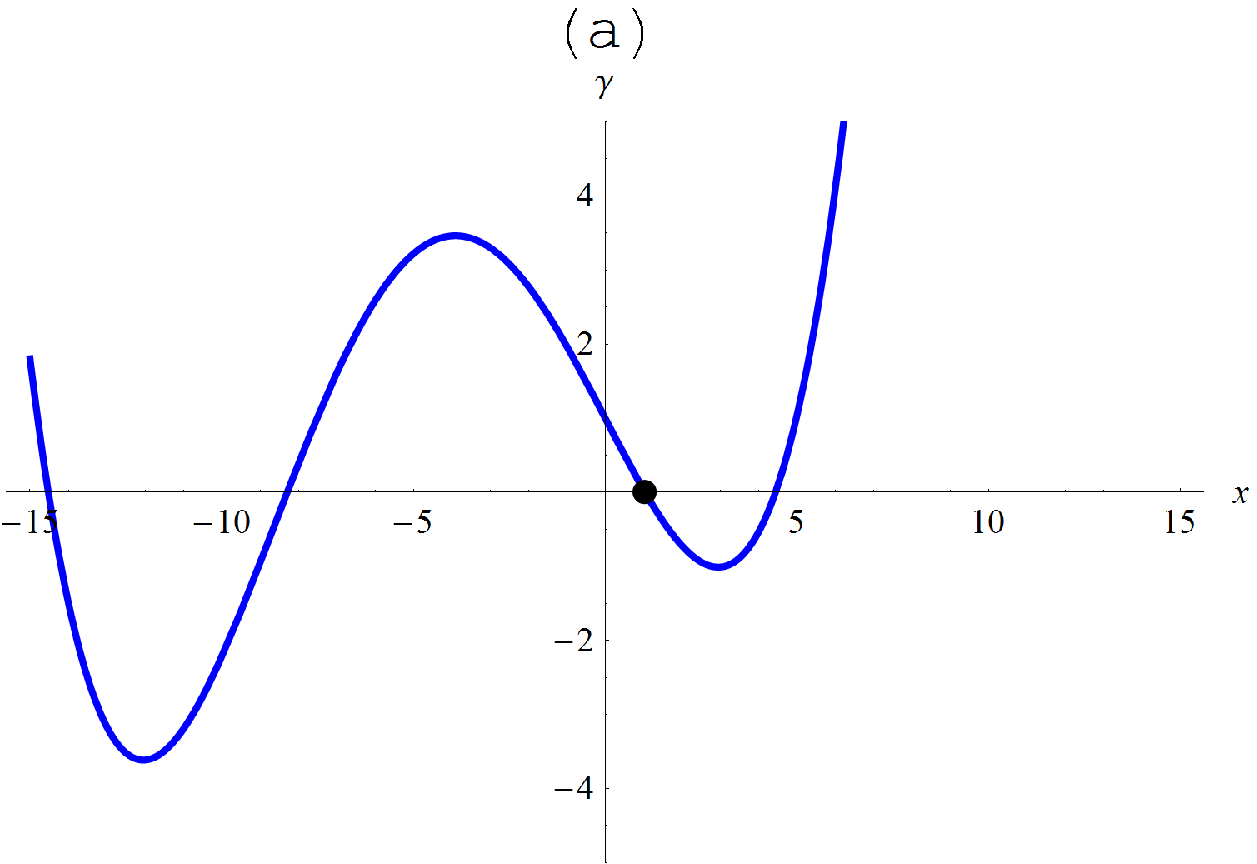} %
\includegraphics[width=0.35\textwidth]{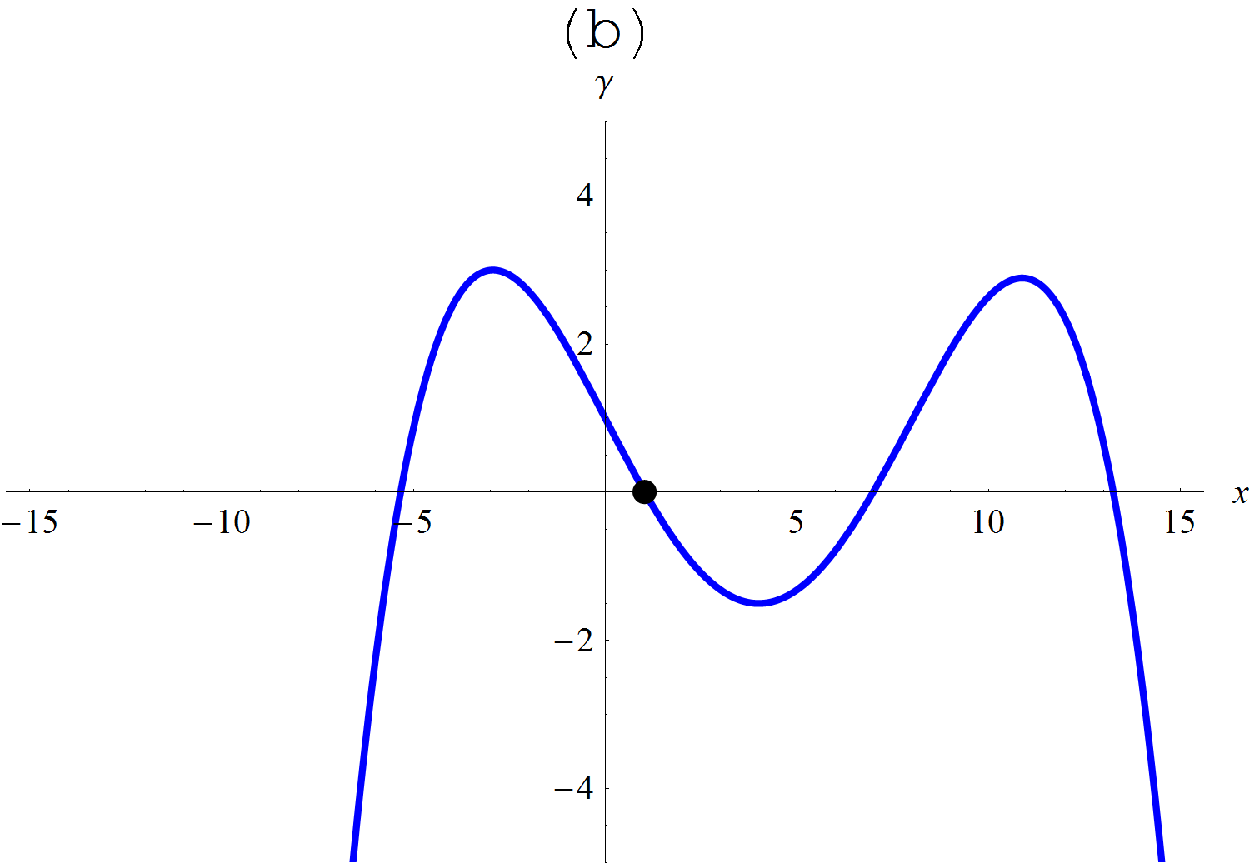} %
\includegraphics[width=0.35\textwidth]{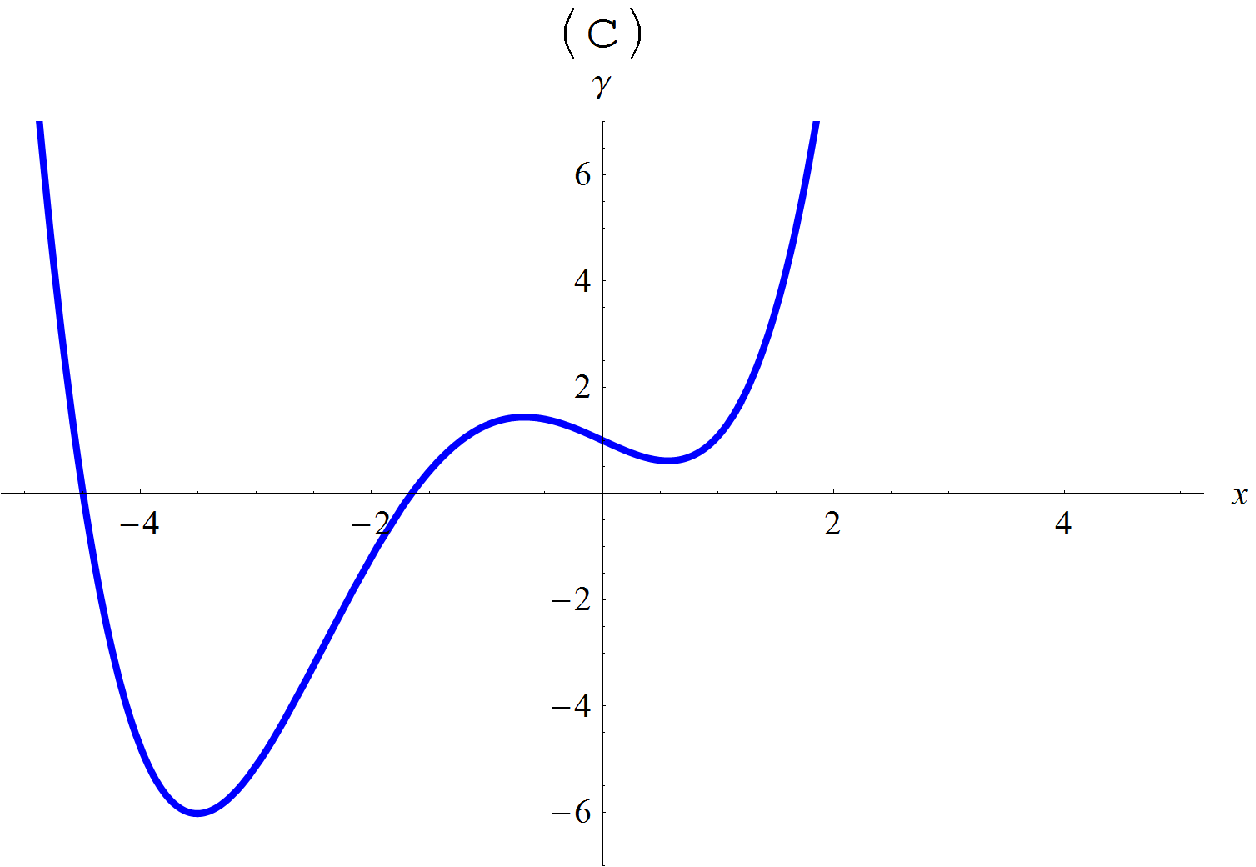} %
\includegraphics[width=0.35\textwidth]{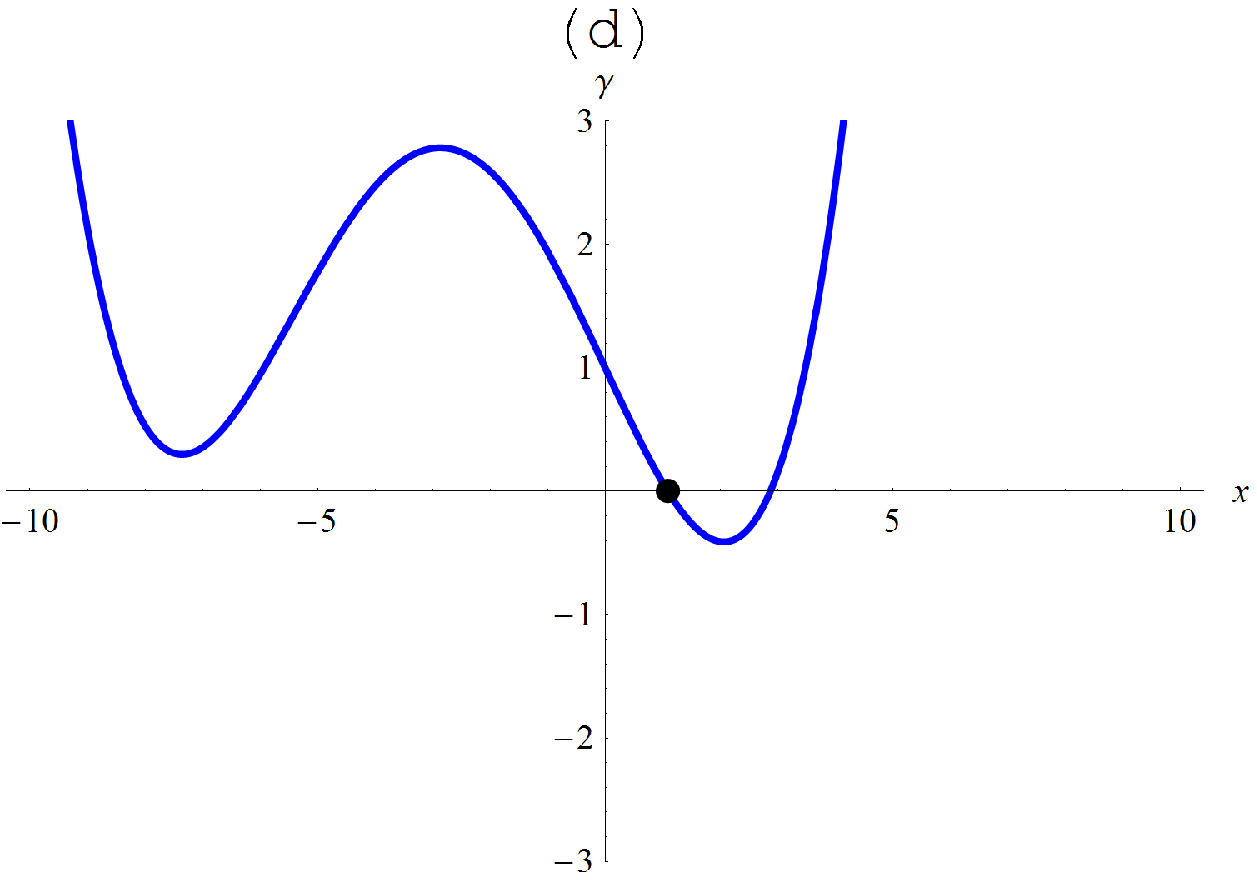} %
\includegraphics[width=0.35\textwidth]{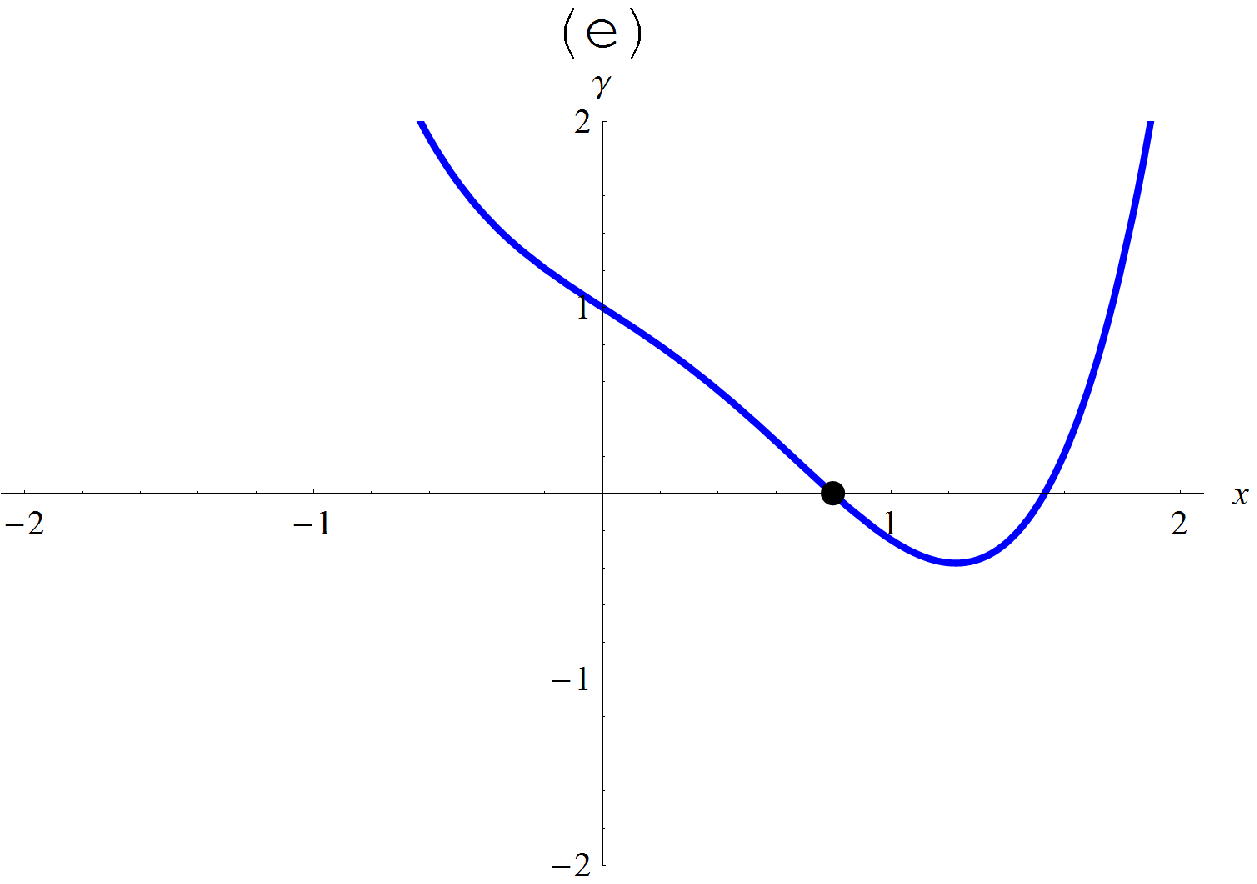} %
\includegraphics[width=0.35\textwidth]{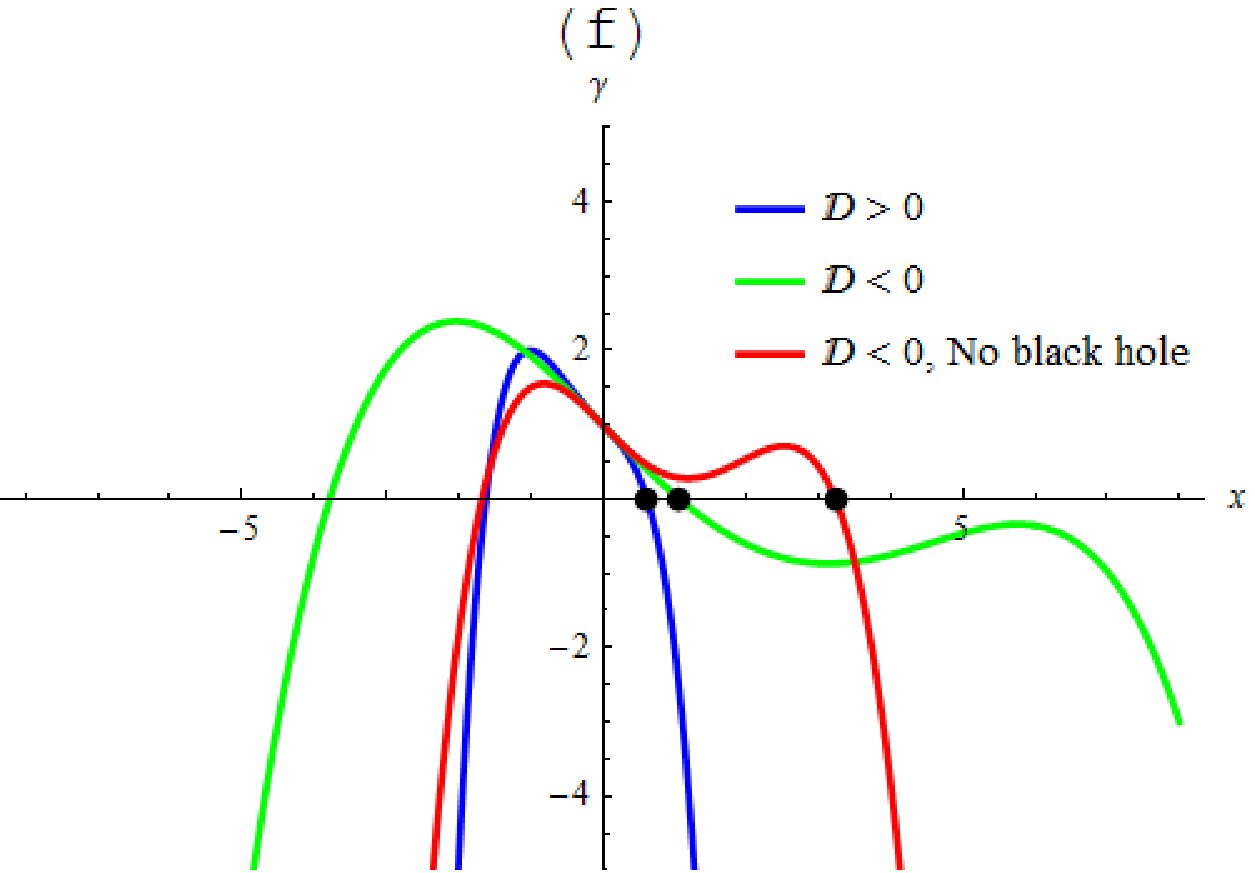}
\caption{Behavior of $\protect\gamma$ as the function of $x$. Black dot in
each figure indicates the ghost free AdS vacuum.}
\label{PAF}
\end{figure}

Consider case (a) of table \ref{Tab1}, which admits four real roots for $%
\gamma =\gamma (x)$ corresponding to two AdS and two dS solutions. This
condition shows the existence of three extrema as one may see in Fig. \ref{PAF}. A positive extremum occur between the two AdS and a negative must
locate between the two dS solutions. Third extremum could be either negative
or positive. However, as we mentioned above the sign of product of the roots
is determined by the sign of $\nu $. So, for region (a) ($\nu >0$) third
extremum must be negative and therefore there is a black hole solution,
which is $f_{3}$. Now let us examine the case (b) of table \ref{Tab1} with
one negative and three positive real roots corresponding to one dS and three
AdS solutions (see Fig. \ref{PAF}). In these cases, there are three extrema.
Two of them occur between the three AdS solutions. But the other one must be
between the smallest AdS solution and the dS one. On the other hand, by
considering Eq. (\ref{extrerma}), this extremum must be negative. So, there
is a black hole solution in this case. It is clear that in this situation
the smallest AdS solution could correspond to asymptotic value of an AdS
planar black hole. In this region $f_{2}$ describes the black hole solution.
In case (c) of the table, there is no AdS solution as it is seen in Fig. \ref%
{PAF}. Next we consider the regions (d) and (e) where there are two AdS
solutions and no dS one. In this cases there is a positive extremum between
the two AdS solutions as one may see in Fig. \ref{PAF}. Since for $\nu >0$
there is no other positive extremum, a black hole solution exists. One may
see the above consideration in Fig. \ref{PAF}.
In the region (f), there are an AdS and a dS solution. An extremum occurs
between the dS and the AdS solutions which may relate to either a black hole
or a naked singularity depending on the sign of this extremum. For $\mathcal{%
D}<0$ there is only one real root for cubic equation which shows the
existence of only one extremum (see Fig. \ref{PAF}). So for $\nu <0$, one
may find that this extremum must locate in the negative region and therefore, a black hole could exist. However for $\mathcal{D}>0$, there are
three real roots for cubic equation and so three extrema which two of them
must be positive. It is possible that one of them locates between zero and
the ghost free AdS. So, this condition excludes black hole solution. But,
one may consider the possibility of existence of black hole for $\mathcal{D}%
>0$ when the two positive roots are greater than the ghost free AdS
solution. One may analyze this situation numerically. The region of the
parameter space for which a black hole may exist has been shown in Fig. \ref%
{Regionf0}. Up to now we study the possibility of existence of any extremum
for $x\in (0,f_{\infty })$. This corresponds to the situation where there is
a real metric function $f(r)$ which admits a horizon at $r_{h}$ and
increases monotonously from $0$ to $f_{\infty }$.

\begin{figure}[tbp]
\centering
\includegraphics[width=0.5\textwidth]{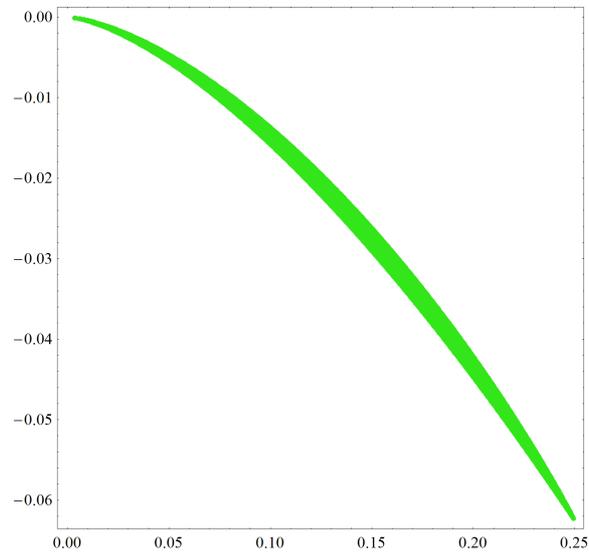}
\caption{This plot displays a part of region (f) for which $\mathcal{D}<0$
and a black hole solution exists. }
\label{Regionf0}
\end{figure}

\begin{table}[th]
\begin{center}
{\footnotesize
\begin{tabular}[h]{|c|c|c|c|c|c|c|c|c|c|}
\hline
& $\Delta $ & $\delta $ & Real Roots & $\nu $ & $\mu $ & dS & Ghost Free AdS
& ghosty AdS & AdS Black Hole \\ \hline
a & + & + & 4 & + & + & 2 & 1 & 1 & $f_{3}$ \\ \hline
b & + & + & 4 & - & + & 1 & 2 & 1 & $f_{2}$ \\ \hline
c & - & $\pm $ & 2 & + & $\mu >0,~G>0$ & 2 & 0 & 0 & - \\ \hline
d & - & $\pm $ & 2 & + & $\mu >0,~G<0$ & 0 & 1 & 1 & $f_{3}$ \\ \hline
e & - & $\pm $ & 2 & + & - & 0 & 1 & 1 & $f_{3}$ \\ \hline
f & - & $\pm $ & 2 & - & $\pm $ & 1 & 1 & 0 &
\begin{tabular}{c}
{\scriptsize $\mathcal{D}>0$, $f_{2}$} \\
{\scriptsize $\mathcal{D}<0$, $f_{2}$ or no black hole}.%
\end{tabular}
\\ \hline
g & + & - & 0 & \multicolumn{6}{|c|}{No Real Roots} \\ \hline
\end{tabular}
}
\end{center}
\caption{Table of various vacua and black hole solutions.}
\label{Tab1}
\end{table}

Before ending this section, we give a comment on the special case $\lambda
=\mu =0$ for which $H$ vanishes. So one may study it with more care. In this
case the discriminant functions $\Delta ,~\delta $, $G$ and $D$ reduce to
\begin{eqnarray*}
\Delta  &=&-\frac{{27}}{{{\nu ^{4}}}}(27+256\nu ),\quad \delta =-\frac{64}{%
\nu } \\
G &=&-\frac{8}{\nu },\quad \mathcal{D}=4\nu ^{4}>0.
\end{eqnarray*}%
Thus, for $\nu >27/256$, $\Delta $ is positive and so there is no real root.
However, for $\nu <27/256$ \ there exist two real roots \cite{Math} which
may be studied as follows. Since $\mathcal{D}>0$, there is only one extremum for $%
\gamma (x)$ which is positive for $\nu <0$ and negative for $\nu >0$ due to
the fact that $\tilde{x}_{1}\tilde{x}_{2}\tilde{x}_{3}=(-1/4\nu )$. Now, by
using asymptotic behaviour of $\gamma (x)\approx \nu x^{4}$ for $%
x\rightarrow \infty $ and $\gamma (0)=1$, it is easy to see that for $\nu >0$
there is a ghost free AdS and a ghosty AdS solution. However, the case $\nu
<0$ admits a ghost free AdS and a dS solution. Thus, for $\nu <27/256$ a
ghost free AdS black hole solution always exists. This solution is given by $%
f_{3}$ for $\nu >0$ and $f_{2}$ for $\nu <0$.

\begin{figure}[tbp]
\centering
\includegraphics[width=0.45\textwidth]{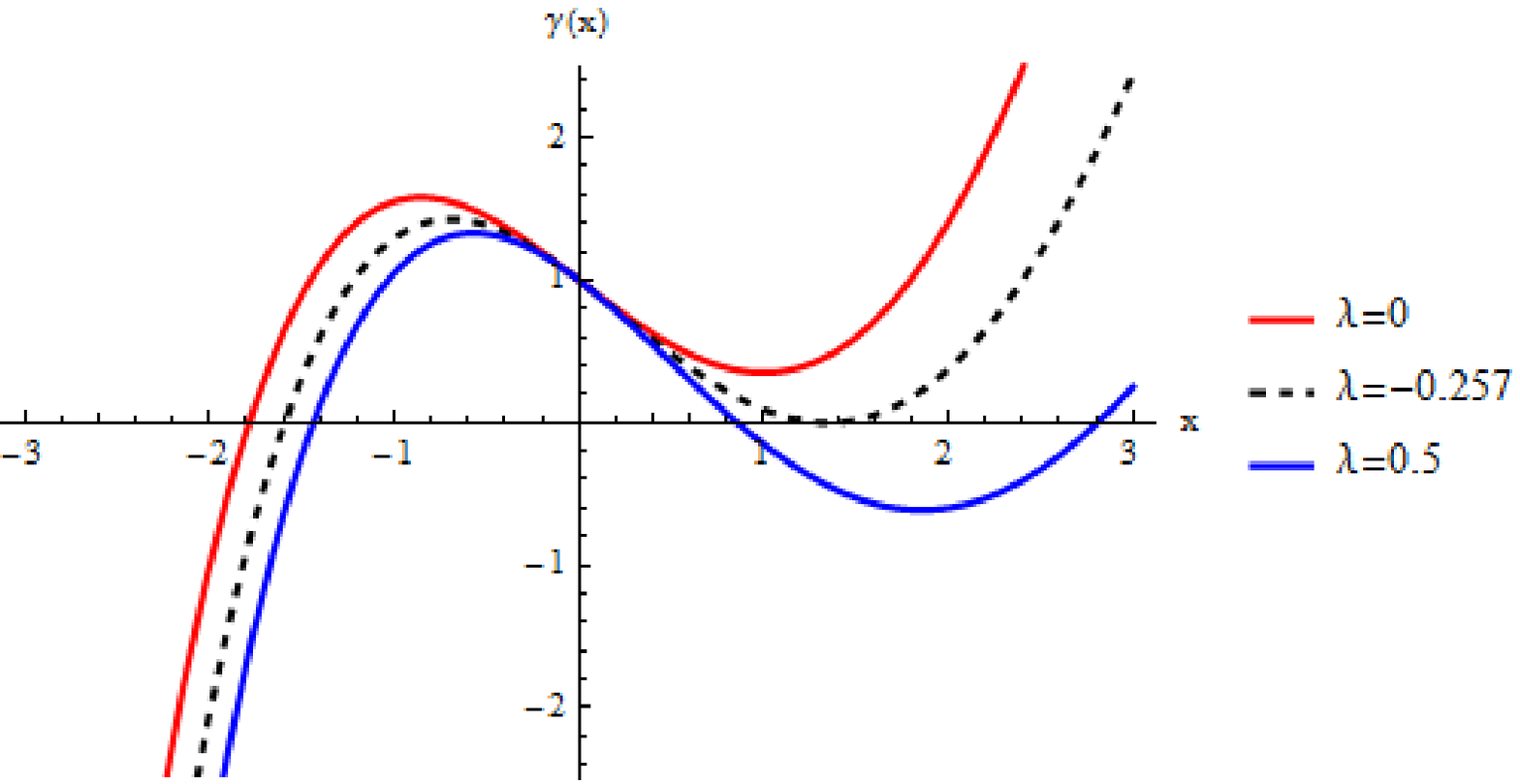} %
\includegraphics[width=0.45\textwidth]{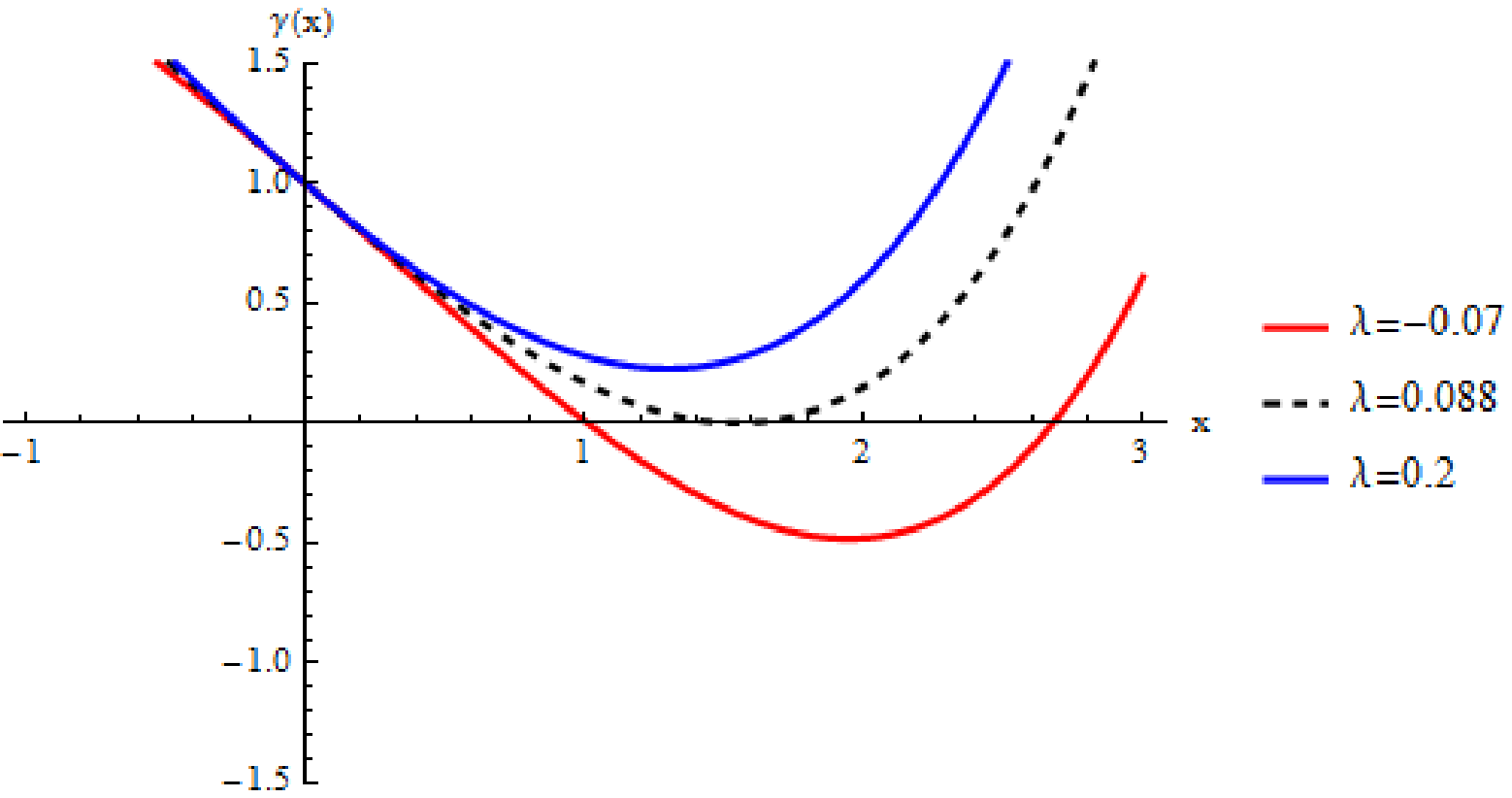} %
\caption{Left figure: $\gamma(x)$ versus $x$ for $\mu=0.4,\nu=-0.05$ and various $\lambda$. Black hole solutions exist for $\lambda<-0.257$.
Right figure: $\gamma(x)$ versus $x$ for $\mu=0.06,\nu=0.02$ and various $\lambda$. Black hole solutions exist for $\lambda<0.088$.}
\label{lambBH}
\end{figure}

Up to now we have considered the cases with $\lambda=0$ and found the regions of parameter space where AdS black holes exist. For investigation of the general cases in the presence of Gauss-Bonnet term, one should use the same argument as the above considerations. We will not study the general case in details, but we consider some special examples which will be used in the following sections. As a first example, we consider a special point in parameter space in a region which does not admit black hole for $\lambda=0$. A point with this property occurs for example at $\mu=0.4$ and $\nu=-0.05$. Figure \ref{lambBH} shows that a black hole solution exists only for $\lambda<-0.257$. As a second example, consider the point ($\lambda=0$, $\mu=0.06$, $\nu=0.02$) in parameter space which is in a region that admits black hole solution. As one may see from Fig. \ref{lambBH}, a black hole exists for $\mu=0.06$, $\nu=0.02$ in the
presence of Gauss-Bonnet term provided $\lambda<0.088$.

\section{Weyl Anomaly and Central Charges\label{Ano}}

In this section, we use the AdS/CFT duality for finding the central charges
of CFT dual to quartic quasi-topological gravity. Indeed, when a CFT is
placed on a curved background the trace of energy-momentum tensor, which is
related to the central charges of CFT, is non-zero and one encounters a Weyl
anomaly (for a historical review see \cite{Duff}). By finding these central
charges, we want to develop the dictionary relating the couplings in
five-dimensional quasi-topological gravity to the parameters which
characterizes its dual four-dimensional CFT. For a CFT in four dimensions,
the trace anomaly relation may be written as
\begin{equation}
\langle T_{a}^{a}\rangle =\frac{c}{16\pi ^{2}}I_{4}-\frac{a}{16\pi ^{2}}%
E_{4},  \label{weyl-anomaly}
\end{equation}%
where
\begin{equation}
I_{4}=R_{ijkl}R^{ijkl}-2R_{ij}R^{ij}+\frac{1}{3}R^{2}\ \ \text{and}\ \ \ \ \
E_{4}=R_{ijkl}R^{ijkl}-4R_{ij}R^{ij}+R^{2}  \label{E4I4}
\end{equation}%
are the square of Weyl tensor and Euler density in four dimensions,
respectively.

In order to compute $a$ and $c$, there is a standard approach in the context
of AdS/CFT correspondence \cite{CentCharg1}. One can start with gravity
action and uses the Fefferman-Graham expansion of the metric
\begin{eqnarray}
ds^{2} &=&\frac{L_{eff}^{2}}{4\rho ^{2}}d\rho ^{2}+\frac{\bar{g}_{ij}}{\rho }%
dx^{i}dx^{j},  \label{Fefferman} \\
\bar{g}_{ij} &=&\bar{g}_{(0)ij}+\rho \bar{g}_{(1)ij}+\rho ^{2}\bar{g}%
_{(2)_{ij}}+...~,
\end{eqnarray}%
where $\bar{g}_{(0)ij}$ denotes the boundary metric. The holographic
approach instructs us to plug this expansion into the action and use the
field equations to eliminate $\bar{g}_{(2)ij}$. The resultant action will be
in terms of $\bar{g}_{(0)ij\text{ }}$ and$~\bar{g}_{(1)ij}$, while the
latter can be found in term of $\bar{g}_{(0)ij\text{ }}$ through the use of
field equations. Then, in order to find the Weyl anomaly, one needs only
those terms which produce a log divergence, that is
\begin{equation}
I\backsim \frac{1}{2}\ln \epsilon ~\int d^{4}x~\sqrt{\bar{g}_{(0)}}\langle
T_{a}^{a}\rangle ,  \label{Lln}
\end{equation}%
where $\epsilon ~$is a cut off limit.

However, doing such messy calculations for higher order gravity are
cumbersome. So we use the nice trick used in Ref. \cite{Sinha} for
calculating the central charges .~We take the boundary $\bar{g}_{ij}$ as
\begin{equation}
d\bar{s}^{2}=u[1+\alpha \rho ](-R^{2}dt^{2}+\frac{dR^{2}}{4R^{2}})+v[1+\beta
\rho ](d\theta ^{2}+\sin ^{2}\theta d\phi ^{2}).  \label{def Ads2S2}
\end{equation}
Thus, the bulk metric is
\begin{equation}
ds^{2}=\frac{L_{eff}^{2}}{4\rho ^{2}}d\rho ^{2}+\frac{1}{\rho }\left\{
u(1+\alpha \rho )\left( -R^{2}dt^{2}+\frac{dR^{2}}{4R^{2}}\right) +v(1+\beta
\rho )(d\theta ^{2}+\sin ^{2}\theta d\phi ^{2})\right\} .  \label{met3}
\end{equation}
We substitute the metric (\ref{met3}) \ into the action and extract the term
proportional to $1/\rho $, which leads to logarithmic divergence. Using the
field equations for the metric (\ref{Fefferman}), and denoting the
logarithmic term as $L_{\ln }$, one finds
\begin{equation}
\frac{\partial }{\partial \alpha }L_{\ln }=0;\ \ \ \ \ \ \ \ \ \ \ \ \ \ \ \
\frac{\partial }{\partial \beta }L_{\ln }=0,  \label{eomalpah-beta}
\end{equation}
which gives $\alpha $ and $\beta $ in terms of other parameters. The
expressions $E_{4}$ and $I_{4}$ for $\bar{g}_{0(ij)}$ background can be
calculated as
\begin{equation}
E_{4}=\frac{-8}{u~v},\ \ \ \ \ \ \ \ \ \ \ \ \ \ \ \ \ I_{4}=\frac{4(u-v)^{3}%
}{3u^{2}v^{2}}.  \label{E4 and I4}
\end{equation}
Now, it is straightforward to show that central charges can be obtained by
using the following equations
\begin{eqnarray*}
c &=&\left( \lim_{v\longrightarrow \infty }\frac{24\pi ^{2}L_{\ln }}{\sqrt{%
\bar{g}_{(0)}}}\right) _{u=1}, \\
a &=&\left( \lim_{v\longrightarrow 1}\frac{4\pi ^{2}L_{\ln }}{\sqrt{\bar{g}%
_{(0)}}}\right) _{u=1}.
\end{eqnarray*}
For quartic quasi-topological gravity such calculations lead to
\begin{eqnarray*}
c &=&\frac{\pi ^{2}L^{3}}{f_{\infty }^{\frac{3}{2}}l_{p}^{3}}(1-2\lambda
f_{\infty }-3\mu f_{\infty }^{2}-4\nu f_{\infty }^{3}), \\
a &=&\frac{\pi ^{2}L^{3}}{f_{\infty }^{\frac{3}{2}}l_{p}^{3}}(1-6\lambda
f_{\infty }+9\mu f_{\infty }^{2}+4\nu f_{\infty }^{3}),
\end{eqnarray*}
which reduce to the central charges of cubic quasi-topological \cite{Myers2}
for $\nu =0$. It is worthwhile to mention that the central charge $c$ is
proportional to the two-points correlation function of energy-momentum
tensor, and therefore in order to have unitary CFT, $c$ should be positive.
This result is in agreement with the condition (\ref{ghost free}) as one may
expect. Indeed both of them arise from the graviton propagation in AdS
background analysis, and therefore they should impose the same constraint on
the parameters of gravity theory.

\section{Causality Constraint in Tensor Channel \label{CC}}

In this section, we perform the causality study of CFT in 4 dimensions duals to
the 5-dimensional quartic quasi-topological gravity. It is known that the
existence of a black hole in the bulk spacetime breaks Lorentz invariant of
dual CFT, and therefore one should consider the possibility of propagation
of superluminal modes in the CFT and violation of causality. In order to
avoid causality violation, one may constrain the gravitational coupling.
Actually, this kind of analysis has been performed for the first time in
Gauss-Bonnet gravity in Ref. \cite{Liu}, and then extended to other higher
order theories of gravity \cite{de Bore1,Edelstein1, de Bore2,GBd,Buchel}.
Here we only study tensor channels analysis, and we leave the other channels analysis for future.

Naturally, the existence of higher curvature terms in a theory of gravity
leads to new coupling parameters and therefore new causality constraints.
However, surprisingly, cubic quasi-topological gravity does not show any new
causality constraints \cite{Myers2}. So the question is:\ does quartic
quasi-topological gravity has the same trivial result? Indeed, one of our
main motivation for studying quartic quasi-topological gravity is these
kinds of investigations. As we will see in this section, considering only
quartic term without the cubic term leads to a trivial result, just like in
the cubic quasi-topological gravity. However, if one considers both cubic and
quartic terms, then a non-trivial result will be appeared. Here we study the
causality in the tensor channel by following the procedure of Ref. \cite{Myers2}%
, closely.

Consider a black hole background such as (\ref{metric}), which is deformed
by a tensor perturbation to
\begin{equation}
ds^{2}=\frac{r_{h}^{2}}{L^{2}\rho }\left( -f(\rho )dt^{2}+d\overrightarrow{x}%
^{2}+2\phi (\rho )e^{i(qx_{3}-\omega t)}dx_{1}dx_{2}\right) +\frac{L^{2}}{%
4\rho ^{2}}\frac{d\rho ^{2}}{f(\rho )},
\end{equation}%
where we define $\rho =r_{h}^{2}/r^{2}$ in Eq. (\ref{f(r)}). The linearized
equation of perturbation $\phi (\rho )$ may be written as
\begin{equation}
\nu \partial _{\rho }[C_{(4)}(\rho )\phi ^{(3)}(\rho )+C_{(4)}^{\prime
}(\rho )\phi ^{^{\prime \prime }}(\rho )]+\partial _{\rho }[C_{(2)}(\rho
,q^{2},\omega ^{2})\phi ^{^{\prime }}(\rho )]+C_{(0)}(\rho ,q^{2},\omega
^{2})\phi (\rho )=0,  \label{phi}
\end{equation}%
where $\phi ^{(i)}$ denotes the $i$th-order derivative of $\phi $ with
respect to $\rho $, $C_{(4)}$ is
\begin{equation}
C_{4}(\rho )=\frac{6464}{73}\rho ^{3}f^{2}r_{h}^{2}(\rho ^{2}f^{\prime
\prime 2}-\frac{22}{101}\rho f^{\prime }f^{\prime \prime }+\frac{3}{4}%
f^{\prime 2}),  \label{C4}
\end{equation}%
and $C_{(0)}$ and $C_{(2)}$ are given in the Appendix. One may note that the
presence of third and fourth-order derivatives of \ $\phi $ with respect to $%
\rho $ is a feature of quartic quasi-topological gravity which is different
from the cubic quasi-topological gravity. We may emphasize that this
perturbation is around a black hole solution and for perturbation around the
AdS solution ($f(\rho )=const.$), $C_{(4)}$ vanishes as one may see in Eq. (%
\ref{C4}). Indeed, in the cubic theory the graviton wave equation around a
black hole background is second order in the radial derivative of $\phi (\rho )$ and
therefore it is convenient to convert it to a Schrodinger like form and
study the graviton wave function. Although in quartic quasi-topological
gravity the graviton wave equation (\ref{phi}) is not second order, but for
our purpose it is sufficient to work at large momentum and frequency limit
of Eq. (\ref{phi}). In fact, the dispersion relation for $\omega $ and $q$
determines the front velocity of the signals as $v_{front}=\lim_{q%
\rightarrow \infty }Re(\omega )/q$. At this limit, we focus on the last term
$C_{(0)}(\rho ,q^{2},\omega ^{2})$, which is the dominant term.

Let's review the case of Gauss-Bonnet gravity ($\mu =0=\nu $), for which the
dominant term reduces to
\begin{eqnarray*}
0 &=&C_{(0)}(\rho ,q^{2},\omega ^{2})=219r_{h}^{2}L^{4}f_{\infty }f(\rho
)(1-2\lambda f(\rho )+2\rho f^{\prime }(\rho ))\omega ^{2} \\
&&-219r_{h}^{2}L^{4}\rho f(\rho )^{2}(1-2\lambda f(\rho )+2\lambda \rho
f^{\prime }(\rho )-4\rho ^{2}\lambda f^{\prime \prime }(\rho ))q^{2},
\end{eqnarray*}
where $f_{\infty }=f(\rho =0)$. By using the near boundary expansion ($%
r\rightarrow \infty $ or $\rho \rightarrow 0$) of $f(\rho )$ in Gauss-Bonnet
gravity, one obtains
\begin{equation}
\frac{\omega ^{2}}{q^{2}}=1-\frac{1-10\lambda f_{\infty }}{(1-2\lambda
f_{\infty })^{2}f_{\infty }}\rho ^{2}+O(\rho ^{4}),
\end{equation}
which shows that the superluminal signals will be avoided, provided $%
1-10\lambda f_{\infty }\geq 0$.

In the quartic quasi-topological gravity, the constant $C_{(0)}(\rho
,q^{2},\omega ^{2})$ contains terms proportional to $q^{2}\omega ^{2}$, $%
\omega ^{4}$ and $q^{4}$ (see Appendix), while in the cubic
quasi-topological gravity there is no $\omega ^{4}$-term. By taking $\omega
^{2}=\alpha ^{2}q^{2}$ and solving for $\alpha ^{2}$, one obtains

\begin{equation}
\lim_{q\rightarrow \infty }\alpha ^{2}=\frac{f}{f_{\infty }}\frac{[1314\rho
f^{\prime \prime }+657f^{\prime }]\mu -[6624\rho ^{3}f^{\prime \prime
2}-24\rho (73f-74\rho f^{\prime })f^{\prime \prime }+168\rho f^{\prime
}{}^{2}-876ff^{\prime }]\nu }{[1314\rho f^{\prime \prime }+657f^{\prime
}]\mu -[14400\rho ^{3}f^{\prime \prime 2}-24\rho (73f-\frac{512}{3}\rho
f^{\prime })f^{\prime \prime }+5708\rho f^{\prime 2}-876ff^{\prime }]\nu }.
\label{alph}
\end{equation}%
It is worth noting that although the above equation does not depend on $%
\lambda $ explicitly, but it depends on $\lambda $ because of the dependence
of $f(r)$ on $\lambda $. For cubic quasi-topological gravity ($\nu =0$),
this equation reduces to $f(\rho )/f_{\infty }$,$~$which is the same as that
of Einstein gravity. But, in quartic quasi-topological gravity, Eq. (\ref%
{alph}) reduces to
\begin{equation*}
\lim_{q\rightarrow \infty }\alpha ^{2}=1-\frac{11300f_{\infty }\nu +657\mu }{%
219(1-2\lambda f_{\infty }-3\mu f_{\infty }^{2}-4\nu f_{\infty
}^{3})(4f_{\infty }\nu +3\mu )f_{\infty }}\rho ^{2}+O(\rho ^{3}),
\end{equation*}%
where we have used the near boundary expansion, $f(\rho )=f_{\infty
}-(1-2\lambda f_{\infty }-3\mu f_{\infty }^{2}-4\nu f_{\infty
}^{3})^{-1}\rho ^{2}+O(\rho ^{3})$. Since $(1-2\lambda f_{\infty }-3\mu
f_{\infty }^{2}-4\nu f_{\infty }^{3})>0$, the following inequality
\begin{equation}
\frac{11300f_{\infty }\nu +657\mu }{(4f_{\infty }\nu +3\mu )f_{\infty }}\geq 0,\label{CAS}
\end{equation}
guarantees the near boundary causality in the tensor channel. It is worth to
note that in the absence of
either of the two quasi-topological terms ($\mu =0$ or $\nu =0$), no
constraint is imposed on the coupling constants because of causality.

\begin{figure}[tbp]
\centering
\includegraphics[width=0.6\textwidth]{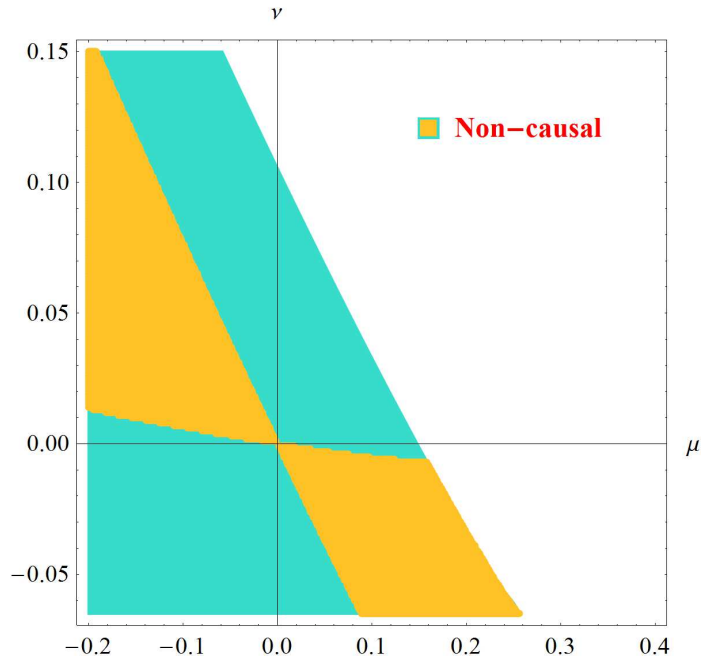}
\caption{The orange and green regions show the non-casual and casual regions, respectively in the tensor channel in the absence of Gauss-Bonnet term.}
\label{causal}
\end{figure}

To determine the region where causality is violated, one need to consider which
roots of $\gamma (f_{\infty })=0$ corresponds to the black hole solution.
Indeed, for each value of \ ($\lambda ,\mu ,\nu $) that admit AdS planar
black hole, we check which root of the quartic equation corresponds to black
hole. Then we take $f_{\infty }$ associate to the black hole root and we plot the region where causality violates in the tensor channel
for the special case $\lambda =0$ (Fig.\ref{causal}). Numerical calculations show that the sensitivity of casual region with respect to variation of $\lambda$ is not considerable. This is due to the fact that Eq. (\ref{CAS}) does not depend on $\lambda$ explicitly and $f_{\infty}$ is of order one.

\section{Holographic hydrodynamics\label{Hydro}}

In order to study the effects of causality on the viscosity/entropy ratio,
we will use the calculations inspired by AdS/CFT correspondence \cite%
{Liu,Kss}. Such calculations have been performed for cubic and quartic
quasi-topological gravity in \cite{Myers2,Dehghani1}. Here, for
completeness, we give a brief review of the interesting pole method \cite%
{Paulos} which has been used in \cite{Dehghani1} in order to obtain shear
viscosity for a planar metric which admits black hole solution. Following
the method of Refs. \cite{Myers2,Paulos} and employing the transformation $%
u=1-r_{h}^{2}/r^{2}$ to map horizon-boundary region $r\in (r_{h},\infty )$
to $u\in (0,1)$, the metric (\ref{metric}) becomes
\begin{equation}
ds^{2}=\frac{r_{h}^{2}}{L^{2}(1-u)}(-\frac{f(u)}{f_{\infty }}%
dt^{2}+dx^{2}+dy^{2}+dz^{2})+\frac{L^{2}}{4f(u)}\frac{du^{2}}{(1-u)^{2}}.
\end{equation}
We expand $f(u)$ around the simple zero at the horizon $u=0$
\begin{equation*}
f(u)=2u+(4\lambda -1)u^{2}+4(2\mu +\lambda (4\lambda -1))u^{3}+O(u^{4}).
\end{equation*}
Now we perturb the metric by the transformation $dx\rightarrow
dx+\varepsilon e^{-i\omega t}dy$, where $\varepsilon ~$is an infinitesimal
positive parameter and evaluate the Lagrangian $\mathcal{\tilde{L}}\equiv
\sqrt{-g}\mathcal{L}_{g}$ on the perturbed metric. The off-shell
perturbation produces a pole at $u=0$ which the residue of this pole gives
us the shear viscosity
\begin{equation}
\eta =8\pi T\lim_{\omega , \rightarrow 0}\frac{\text{Res}_{u=0}\mathcal{L}}{%
\omega ^{2}\varepsilon ^{2}}.  \label{Res}
\end{equation}
In Eq. (\ref{Res}) $T$ is the temperature of black hole given in Eq. (\ref%
{Temper}) and Res$_{u=0}\mathcal{\tilde{L}~}$ stands for the residue of the
pole at $u=0$. One obtains the viscosity/entropy ratio as
\begin{eqnarray}
\frac{\eta }{s} &=&\frac{1}{4\pi }[1-4\lambda -36\mu (9-64\lambda
+128\lambda ^{2}+48\mu )  \notag \\
&&-\frac{96}{73}\nu (1491+6240\mu -10752\mu \lambda  \notag \\
&&-10800\lambda +28864\lambda ^{2}-25088\lambda ^{3})],
\end{eqnarray}
\begin{figure}[tb]
\centering
\includegraphics[width=0.6\textwidth]{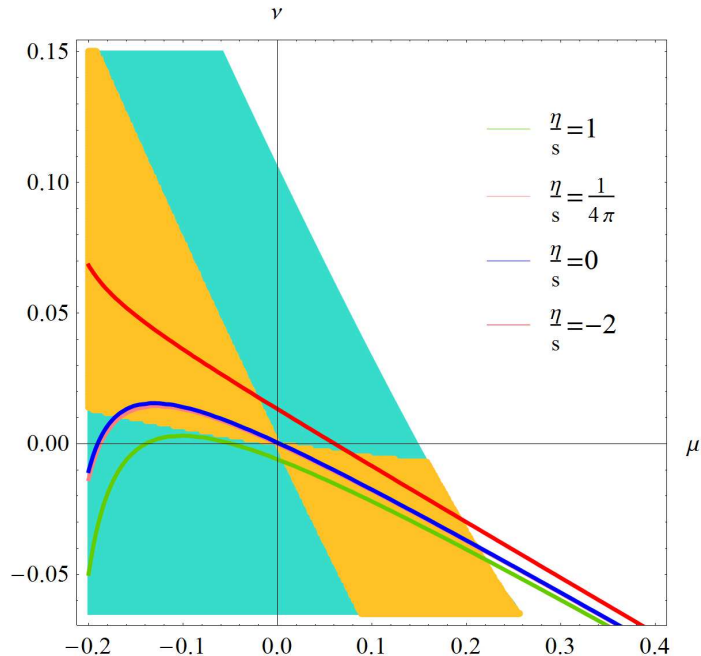}
\caption{Contours of constant $\frac{\protect\eta }{s}$}
\label{etas}
\end{figure}
which shows that, in contrast to Lovelock gravity, the cubic and quartic
terms have contributions in the $\eta /s$ \cite{shearlovelock}. For $\lambda
=0$ this ratio reduce to
\begin{equation}
\frac{\eta }{s}=\frac{1}{4\pi }[1-36\mu (9+48\mu )-\frac{96}{73}\nu
(1491+6240\mu )].  \notag
\end{equation}

In Fig. \ref{etas} we plot the contours $\eta /s=const$ on the ($\mu ,\nu )$
plane. From this plot, it is clear that the near boundary causality
constraint in the tensor channel is not strong enough to preserve a positive bound on $\eta /s$. Of course, this fact is a common feature of higher order curvature
gravities such as third order Lovelock or cubic quasi-topological gravities
\cite{Edelstein1,de Bore2}. However in the case of cubic quasi-topological
gravity, the positive energy constraint restores a lower positive bound on $%
\eta /s$ which excludes the negative value for $\eta /s$ \cite{Myers2}.
Thus, we expect that the positive energy constraints in different channels
may restore a positive bound on $\eta /s$ in our case too, but we leave it
for future.

\begin{figure}[tb]
\centering
\includegraphics[width=0.6\textwidth]{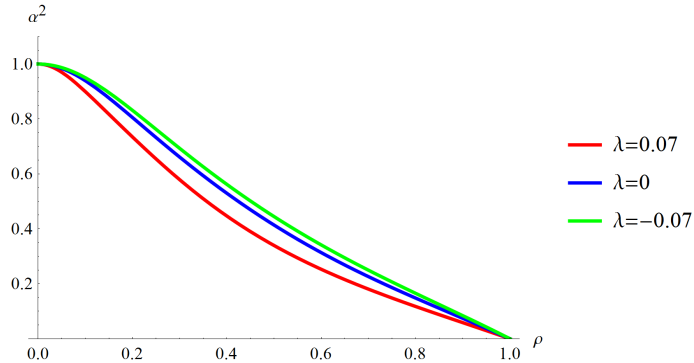}
\caption{Plot of $\protect\alpha ^{2}$ in term of $\protect\rho $ for $%
\protect\mu =.06$ and $\protect\nu =.02$. The plot shows the $\protect\alpha %
^{2}$ is positve and less than one for every where between horizon and
boundary.}
\label{BulkCaus}
\end{figure}

Another possibility is the consideration of interior geometry
causality constraint. We studied the tensor channel causality in the near
boundary by using the asymptotic expansion of the metric function and found some
constraints on the coupling constants. However, as mentioned in \cite%
{EdelsteinPaulos}, the causality may be violated in the interior geometry in
Lovelock gravity. So, studying the causality in the bulk may imply a
stronger constraint on the parameter space and perhaps restores a positive
bound on $\eta /s$. General studies of this issue is cumbersome, but to
understand the probable effects of bulk causality on the negativity of $\eta
/s$ we take $\mu =.06$, $\nu =.02$ and $\lambda =-0.07,0,0.07$. As one can see in Fig. \ref{lambBH},
for these choices a black hole solution exists.  In Fig.\ref{BulkCaus}, we plot $\alpha ^{2}$ by
substituting the metric function of the black hole solution in Eq. (\ref{alph}). This plot shows that $%
0\leq \alpha ^{2}\leq 1$. Also, these values lead to negative viscosity.
Thus, it seems that this constraint has no chance to keep $\eta /s$ to be
positive. However, there are some restrictions about our result. Indeed,
the existence of higher derivative terms in the field equation of graviton
on the black hole background does not allow us to write the wave equation
in the neat Schrodinger form and analyze the potential. So, perhaps one
needs to treat the effective dispersion relation $\alpha $ with more care.
Also, it is worth to mention that our investigation is done for 5-dimensional spacetime and the bulk causality constraint is sensitive to the dimension of the spacetime.
In addition, there is another source of constraints for the
gravitational coupling coming from the instabilities of the plasma in the
dual theory \cite{GBd}. Indeed, this issue occurs when the local speed of
graviton becomes imaginary $\alpha ^{2}<0$. This constraint,
implies a lower positive bound on $\eta /s$ in Lovelock theory
\cite{EdelsteinPaulos}, but as we have seen above this cannot implies any
positive lower bound on $\eta /s$ in 5-dimensional quasi-topological gravity. But, as we have mentioned above
this investigation is sensitive to the dimension of the spacetime.

\section{Concluding Remarks \label{ConR}}
In this paper, we considered quartic quasi-topological gravity in the
presence of negative cosmological constant in five dimensions. We varied the action with
respect to the metric $g^{ab}$ and found the general tensorial form of field equation for
quartic quasi-topological gravity with cubic and quartic terms in Riemann
tensor. In general, this field equation includes forth-order derivatives.
But for the special choices of coefficients, $c_{i}$'s, given in Eq. (\ref{cis}) and spherically symmetric
spacetimes, the field equation is second-order. Moreover, we showed that a
linearized perturbation around an AdS spacetime reduces to a second-order
wave equation for graviton which matches to Einstein's gravity up to an
overall factor. By considering the sign of this overall factor, we
constrained the coupling parameters in order to have ghost free AdS solutions.
In the case of quasi-topological gravity (as like as Lovelock gravity), the
field equation for spherically symmetric ansatz leads to a polynomial
equation (quartic equation for quartic quasi-topological gravity). For
simplicity, we analyzed the quartic equation in the absence of Gauss-Bonnet
term and found the region of parameter space where ghost free AdS solutions
exist. In addition, we determined the region of parameter space where the
existence of planar AdS black hole is possible. As one may expect, there are
some regions where ghost free AdS solution exists but there is no black
hole. These cases occur when there are double roots solutions for the quartic
equation. In these cases, by studying the Kretchman curvature scalar, we
explicitly showed that the spacetime involves a naked singularity. However,
it is interesting to generalize these discussions for curved boundary and
higher-dimensional spacetime \cite{Lovebes}. We, also, considered
the general case in the presence of Gauss-Bonnet term for a few special values of $\mu$ and $\nu$.

In the context of AdS/CFT, we found the central charges of CFT$_{4}$ dual to
the quartic quasi-topological gravity. We found that the unitary of the CFT$_{4}$%
, which\ implies the positivity of central charge and relates to the
energy-momentum tensor two-point function, is the same as the stability
constraint which comes from the study of perturbation around AdS$_{5}$. We
also studied tensor-channel perturbation around a black hole background to
find the effects of quasi-topological gravity on the causality violation. 
We obtained the equation for linearized perturbation around a black
hole solution and found that it includes forth-order derivatives. However,
in the absence of the quartic term this equation reduces to a second-order
one. Moreover, in contrast to cubic quasi-topological and (Weyl)$^{2}$ gravities which
do not show any causality violation, we showed that the existence of cubic
and quartic terms together provide the possibility of causality violation.
So in order to survive causality, one needs to imply some constraints on the
region of parameter space where black holes may exist. We also considered
causality constraints in the tensor channel on the viscosity/entropy ratio and found 
that this constraint is not strong enough to imply any positive bound on the viscosity/entropy
ratio. We also studied the possibility of violation of tensor channel causality and instabilities 
in the interior geometry in five dimensions for some special cases. But, the positivity of $\alpha^2$ 
does not imply any positive bound on $\eta/s$. Although this is in contrast to Lovelock gravity in seven dimensions,
but one should note that $\alpha^2$ is sensitive to the dimensions of the spacetime. Also, because of 
the existence of higher order derivatives, perhaps more careful study of quasi normal modes is necessary.

It is known that for supersymmetric theory the causality and positive energy
conditions are related \cite{Hofman}. Here, we address the interesting study
in relation between causality and positive energy constraints for a
non-supersymmetric theory like quasi-topological gravity \cite{Myers2,
Hofman}. Indeed, this was one of the essential motivations for studying
quasi-topological gravity in a five-dimensional spacetime. But this comparison
is not possible in cubic gravity frame work, since the causality does not
imply any constraint \cite{Myers2}. However in quartic quasi-topological
gravity, as we have discussed above, the causality in the tensor channel creates a new constraint
but it is not strong enough to imply any bound on the viscosity/entropy
ratio. Thus, one may consider the possibility of causality violation in other channels \cite{GBd} or positive energy constraint for
quartic quasi-topological gravity in order to survive this ratio. In addition, the full study of perturbation
around a black hole solution in the various channels \cite{GBd, othercha} and finding the
quasi-normal modes in order to find the possibility of causality violation or 
instability are interesting. They may imply a positive bound on $\eta/s$ as in the case of Lovelock theory \cite{GBd,EdelsteinPaulos}. Besides,
as it is mentioned in \cite{EnerUnit} the positivity of energy flux is
equivalent to the absence of ghost in the finite temperature conformal field
theory. So, it is worthwhile to study this property directly in the
quasi-topological gravity. Also by using the central charges
calculated in this paper, one may study the effects of quartic
quasi-topological term on the holographic c-theorem \cite{Ctheo} and
entanglement entropy\cite{Ryu}. We leave these interesting topics for future
works.
\acknowledgments
We are grateful to the referee for constructive comments which helped us to improve the paper significantly. 
M. H. Vahidinia would like to thank S. Jalali, S. Zarepour and A. Naseh for
useful discussions. This work has been supported financially by Research
Institute for Astronomy and Astrophysics of Maragha (RIAAM), Iran.

\appendix

\section{The coefficients $C_{(0)}$ and $C_{(2)}$}

The constant $C_{(0)}(\rho ,q^{2},q^{4},\omega ^{2},\omega ^{4})=-\tilde{C}%
_{(0)}/(438r^{3}f^{2}r_{h}^{2}L^{4})$ in Eq. (\ref{phi}) for the case of
quartic quasi-topological gravity is
\begin{eqnarray*}
\tilde{C}_{(0)} &=&219\,{L}^{4}{f_{\infty }}r_{h}^{2}\rho f[1-2\lambda
(f+\rho {f^{\prime }}\,)]{\omega }^{2}+219\,{L}^{4}r_{h}^{2}\rho {f}%
^{2}[-1+2\,f\lambda +4\,{\rho }^{2}f^{\prime \prime }\,\lambda -2\,\rho
f^{\prime }\,\lambda ]{q}^{2} \\
&&-657\,{L}^{4}f_{\infty }{r}_{h}^{2}\,\rho f[12\,{\rho }^{4}f^{(4)}f+6\,{%
\rho }^{3}\left( \rho f^{\prime }+8\,f\right) f^{(3)}+6\,{\rho }%
^{4}f^{\prime \prime 2}+\left( 18\,{\rho }^{3}f^{\prime }+27\,{\rho }%
^{2}f\right) f^{\prime \prime } \\
&&+{f}^{2}+3\,{\rho }^{2}f^{\prime 2}-2\,\rho f^{\prime }\,f]\mu \,\omega
^{2}-12\,{L}^{4}f_{\infty }{r}_{h}^{2}\rho \,\{\left( 944\,{\rho }%
^{6}f^{\prime \prime }\,-1064\,{\rho }^{5}f^{\prime }\,+876\,{\rho }^{4}{f}%
\right) {f}^{2}f^{(4)} \\
&&+944\,{\rho }^{6}f^{(3)2}{f}^{2}+((-1952\,{\rho }^{6}f^{\prime }\,+4872\,{%
\rho }^{5}{f})f^{\prime \prime }+3504\,{\rho }^{3}{f}^{2}-4042\,{\rho }%
^{4}f^{\prime }\,{f} \\
&&+[(-1952\,{\rho }^{6}f^{\prime }\,+4872\,{\rho }^{5}{f})f^{\prime \prime
}+3504\,{\rho }^{3}{f}^{2}-4042\,{\rho }^{4}f^{\prime }\,{f}-268\,{\rho }%
^{5}f^{\prime 2}]ff^{(3)} \\
&&-976\,{\rho }^{6}f^{\prime \prime 3}f+\left( 808\,{\rho }^{6}f^{\prime
2}+1514\,{\rho }^{4}{f}^{2}-4440\,{\rho }^{5}f^{\prime }\,f\right) f^{\prime
\prime 2} \\
&&+\left( 1971\,{f}^{3}{\rho }^{2}-1234\,{f}^{2}{\rho }^{3}f^{\prime }-176\,{%
\rho }^{5}f^{\prime 3}-3426\,{\rho }^{4}ff^{\prime 2}\right) f^{\prime
\prime } \\
&&+606\,{\rho }^{4}f^{\prime 4}-219\,{f}^{3}f^{\prime }\,\rho +73\,{f}%
^{4}-1821\,{\rho }^{3}f^{\prime 3}f+452\,{f}^{2}f^{\prime 2}{\rho }^{2}\}\nu
\omega ^{2} \\
&&-657\,\,{L}^{4}{\ }r_{h}^{2}\rho {f}^{2}[12{\rho }^{4}f\ f^{(4)}+12{\rho }%
^{3}\left( \rho f^{\prime }+4\,f\right) f^{(3)}+(23\,{\rho }^{2}f+24\,{\rho }%
^{3}f^{\prime })f^{\prime \prime }-\,(\rho f^{\prime }-f)^{2}]\mu {q}^{2} \\
&&+12\,{L}^{4}r_{h}^{2}\rho {f}^{2}[\left( 1368\,{\rho }^{5}f^{\prime
}\,f-592\,{\rho }^{6}f^{\prime \prime }\,f-876\,{\rho }^{4}{f}^{2}\right)
f^{(4)}-592\,{\rho }^{6}f^{(3)2}f \\
&&+((-592\,{\rho }^{6}f^{\prime }+1280\,{\rho }^{5}f)f^{\prime \prime
}+5428\,{\rho }^{4}f^{\prime }\,f-3504\,{f}^{2}{\rho }^{3}+1368\,{\rho }%
^{5}f^{\prime 2})f^{(3)} \\
&&+\left( 1140\,{\rho }^{5}f^{\prime }+6496\,{\rho }^{4}f\right) f^{\prime
\prime 2}+(1334\,ff^{\prime }\,{\rho }^{3}+3076\,\rho ^{4}f^{\prime 2}-1533\,%
{f}^{2}{\rho }^{2})f^{\prime \prime } \\
&&+73\,{f}^{3}-657\,{\rho }^{3}f^{\prime 3}-219\,f^{2}f^{\prime }\,\rho
-219\,ff^{\prime 2}\rho ^{2}]\nu {q}^{2} \\
&&-6\,{L}^{8}f_{\infty }^{2}{\rho }^{4}(-88\,\rho {\ }f^{\prime }\,f^{\prime
\prime }+303\,f^{\prime 2}+404\,{\rho }^{2}f^{\prime \prime 2})\nu \,\omega
^{4}+657\,{L}^{8}{\rho }^{3}{f}^{2}\left( f^{\prime }+2\,\rho f^{\prime
\prime }\right) \mu {q}^{4} \\
&&-3\,{L}^{8}{\rho }^{3}{f}^{2}\left( 2208\,{\rho }^{3}f^{\prime \prime
2}-8\,\rho \left( -74\,\rho f^{\prime }+73\,f\right) f^{\prime \prime
}+56\,\rho f^{\prime 2}-292\,f^{\prime }\,f\right) \nu {q}^{4} \\
&&+{L}^{8}f_{\infty }{\rho }^{3}f\,(-657\,f^{\prime }-1314\,\rho ~f^{\prime
\prime })\mu {q}^{2}\omega ^{2} \\
&&+{L}^{8}f_{\infty }\,{\rho }^{3}f(14400\,{\rho }^{3}f^{\prime \prime
2}-8\,\left( -512\,\rho f^{\prime }+219\,f\right) \rho f^{\prime \prime
}+5708\,\rho f^{\prime 2}-876\,f^{\prime }\,f)\nu {q}^{2}\omega ^{2}.
\end{eqnarray*}
which contains terms proportional to $q^{2}\omega ^{2}$, $\omega ^{4}$ and $%
q^{4}$. Also, $C_{(2)}$ is
\begin{eqnarray*}
C_{(2)} &=&[6\,f(\rho )\rho \,(2\,\rho \,{f^{\prime \prime }}+{f^{\prime }}%
)\mu +{\frac{8}{219}}\,\rho \,f(438\,\rho \,{f^{\prime \prime }}\,f+219\,{%
f^{\prime }}\,f \\
&&-1024\,{\rho }^{2}{f^{\prime \prime }}\,{f^{\prime }}-3600\,{\rho }^{3}{{%
f^{\prime \prime }}}^{2}-1427\,\rho \,{{f^{\prime }}}^{2})\nu ]{q}^{2} \\
&&+{\frac{8}{73}}\,{\rho }^{2}{f_{\infty }}\,[-88\,\rho {f^{\prime \prime }}%
\,{f^{\prime }}+404\,{\rho }^{2}{{f^{\prime \prime }}}^{2}+303\,{{f^{\prime }%
}}^{2}]\nu \,{\omega }^{2} \\
&&+6\,{\frac{f{r}_{h}^{2}\mu }{\rho \,{L}^{4}}}[6\,{f^{\prime }}\,{\rho }^{4}%
{f^{(3)}}+18\,{f^{\prime }}\,{\rho }^{3}{f^{\prime \prime }}+6\,{\rho }^{4}{{%
f^{\prime \prime }}}^{2}+3\,{\rho }^{2}{{f^{\prime }}}^{2}+f^{2}-2\,\rho \,f{%
f^{\prime }}] \\
&&+{\frac{4}{73}}\,{\frac{f{r0}^{2}\nu }{\rho \,{L}^{4}}}[((2560\,{\rho }^{6}%
{f^{\prime }}-352\,{\rho }^{5}f){f^{\prime \prime }}-1240\,{\rho }^{5}{{%
f^{\prime }}}^{2}+572\,f{f^{\prime }\rho }^{4}){f^{(3)}}+1280\,{\rho }^{6}{{%
f^{\prime \prime }}}^{3} \\
&&+(4256\,{\rho }^{5}{f^{\prime }}-1344\,f{\rho }^{4}){{f^{\prime \prime }}}%
^{2}+(1716\,f{f^{\prime }\rho }^{3}-2044\,{\rho }^{4}{{f^{\prime }}}^{2}){%
f^{\prime \prime }}-438\,\rho \,f^{2}{f^{\prime }} \\
&&+1081\,{\rho }^{2}{{f^{\prime }}}^{2}f+146\,f^{3}+2418\,{\rho }^{3}{{%
f^{\prime }}}^{3}]-\,{\frac{2{r}_{h}^{2}f}{\rho \,{L}^{4}}}(1-2\,\lambda
(\rho \,\,{f^{\prime }}-\,f)).
\end{eqnarray*}

\end{document}